\documentclass[aps,showpacs,showkeys,amsmath,amssymb,nofootinbib,pra,floatfix,preprint]{revtex4}

\usepackage{graphicx}
\usepackage{dcolumn}
\usepackage{slashed}
\usepackage{bm}
\usepackage{xcolor}

\usepackage{amssymb}
\usepackage{amsmath}

\usepackage{psfrag,graphicx} 

\newcommand{\sech}{{\rm sech \,}}

\newcommand{\be}{\begin{equation}}
\newcommand{\ee}{\end{equation}}
\newcommand{\bea}{\begin{eqnarray}}
\newcommand{\eea}{\end{eqnarray}}

\newcommand{\vc}[1]{\mbox{\boldmath$#1$}}

\newcommand{\ssvc}[1]{\mbox{\scriptsize\boldmath$#1$}}
\newcommand{\vcb}[1]{\mbox{\bf #1}}

\newcommand{\Ep}{E_{\ssvc{p}}}

\begin{document}
\title{Beam-Shape Effects in Nonlinear Compton and Thomson Scattering}


\author{T.~Heinzl$^1$, D.~Seipt$^2$ and B.~K\"ampfer$^2$}
\affiliation{$^1$School of Computing and Mathematics, University of Plymouth, Drake Circus, Plymouth PL4 8AA, UK\\
$^2$Forschungszentrum Dresden-Rossendorf, POB 51 01 19, 01314 Dresden, Germany}

\date{\today}


\begin{abstract}
We discuss intensity effects in collisions between beams of optical photons from a high-power laser and relativistic electrons. Our main focus are the modifications of the emission spectra due to realistic finite-beam geometries. By carefully analysing the classical limit we precisely quantify the distinction between strong-field QED Compton scattering and classical Thomson scattering. A purely classical, but fully covariant, calculation of the bremsstrahlung emitted by an electron in a plane wave laser field yields radiation into harmonics, as expected. This result is generalised to pulses of finite duration and explains the appearance of line broadening and harmonic substructure as an interference phenomenon. The ensuing numerical treatment confirms that strong focussing of the laser leads to a broad continuum while higher harmonics become visible only at moderate focussing, hence lower intensity. We present a scaling law for the backscattered photon spectral density which facilitates averaging over electron beam phase space. Finally, we propose a set of realistic parameters such that the observation of intensity induced spectral red-shift, higher harmonics, and their substructure, becomes feasible.
\end{abstract}

\pacs{12.20.Ds,41.60.-m,}
\keywords{high-intensity lasers, Compton scattering, Thomson scattering}

\maketitle

\section{Introduction}

Intensity (or strong-field) effects on QED scattering processes have been investigated since the 1960s following the invention of the laser. The pioneering studies considered both strong-field pair creation \cite{Reiss:1962} and the crossed process, electron photon scattering \cite{Nikishov:1963,Nikishov:1964a,Nikishov:1964b,Goldman:1964,Narozhnyi:1964,Brown:1964zz,Kibble:1965zz} where the use of laser beams has already been suggested. Since then there has been a wealth of theoretical papers and we refer the reader to the reviews \cite{McDonald:1986zz,Fernow,Lau:2003,Mourou:2006zz,Salamin:2006ff,Marklund:2006my} for an overview of the literature relevant for the present subject. From an experimental point of view the situation is less straightforward. There have only been a few clear-cut observations of intensity dependent effects. Probably the best known experiment is SLAC E-144 probing strong-field QED using a (by now moderately) intense laser beam in conjunction with high-energy electrons \cite{Bamber:1999zt}. Colliding a laser $(L)$ of intensity $10^{18}$ W/cm$^2$ with the $46.6$ GeV electron beam the observation of the nonlinear Compton scattering process
\begin{equation}
  e + \ell \gamma_L \rightarrow  e' + \gamma
  \label{reaction.nlcompton}
\end{equation}
has been reported in \cite{Bula:1996st}. Note that without an external field (here provided by the laser) an electron could not spontaneously emit photons as this is forbidden by energy momentum conservation. However, absorption of $\ell$ laser photons $\gamma_L$ induces the production of a high-energy ($30\ \rm GeV$) $\gamma$ quantum which thus takes away a large fraction of the incoming electron energy. This high-energy photon has then been used to produce electron-positron pairs \cite{Burke:1997ew} via collision with the laser, employing the multi-photon Breit-Wheeler reaction \cite{Breit:1934}, $\gamma + \ell^\prime \gamma_L \rightarrow e^+ + e^{-}$. Hence, using a high-energy setting with a large linac, SLAC E-144 has produced ``matter from light'' for the first time \cite{Burke:1997ew}.

It is convenient to describe electron energy and laser intensity in terms of dimensionless parameters. The former is of course measured in terms of the relativistic gamma factor, $\gamma = \Ep/m \equiv \cosh \zeta$ where $\zeta$ denotes rapidity. A convenient measure of laser intensity is the dimensionless laser amplitude,
\begin{equation}  \label{A0}
  a_0 \equiv \frac{e E \lambdabar}{m} = \frac{e E}{m\omega} \equiv \frac{e E}{m^2\nu}\; ,
\end{equation}
with $E$ being the root-mean-square electric field and $\lambdabar = 1/\omega = 1/m\nu$ the laser wavelength. $a_0$ is thus a purely classical quantity, the energy gain of an electron across a wavelength measured in units of its rest mass. We mention in passing that quantum field theory imposes an upper limit on $a_0$ \cite{Bulanov:2004de} which becomes manifest upon writing
\be
  a_0 = \frac{e E}{m^2} \frac{1}{\nu} \equiv \frac{E}{E_S} \frac{1}{\nu} \; .
\ee
Here $E_S = m^2/e = 1.3 \times 10^{18}$ V/m is the QED critical field first discussed by Sauter \cite{Sauter:1931zz} beyond which any laser becomes unstable due to pair creation from the vacuum via the Schwinger mechanism \cite{Schwinger:1951nm}. For an optical laser this implies the bound $a_0 < 1/\nu \simeq 10^6$.

The definition (\ref{A0}) can be made explicitly Lorentz and gauge invariant \cite{Heinzl:2008rh}. When $a_0$ becomes of order unity the quiver motion of the electron in the laser beam becomes relativistic. SLAC E-144, for instance, had $a_0 \simeq 0.6$ and $\gamma = 10^5$, i.e.\ low intensity and high energy for the purposes of this paper.

From a classical (nonlinear optics) point of view the process (\ref{reaction.nlcompton}) corresponds to the generation of the $\ell$th harmonic in the $\gamma$ radiation spectrum. The production of higher harmonics has been observed in several experiments colliding laser and electron beams: low intensity laser photons ($a_0 = 0.01$) with low-energy ( $\sim$ 1 keV) electrons from an electron gun \cite{Englert:1983zz}, $a_0 = 2$ photons with plasma electrons from a gas jet \cite{Chen:1998} and, more recently, sub-terawatt laser photons ($a_0 = 0.35$) with $60$ MeV electrons from a linac \cite{Babzien:2006zz}.  Using linearly polarised photons the latter two papers have analysed the characteristic azimuthal intensity distributions confirming quadrupole and sextupole patterns for the second and third harmonics, respectively. The energy spectrum of the scattered radiation, to the best of our knowledge, has been measured only once, in an all-optical setup using laser accelerated electrons \cite{schwoerer:2006}. While this ``table-top" setup is certainly attractive as it does not require a linac, the electron beam has a rather broad and random energy distribution which in turn is inherited by the scattered photons. As a result, the $\gamma$ spectrum recorded in \cite{schwoerer:2006} is rather difficult to analyse theoretically.

In this paper we discuss the prospects of experimentally analysing nonlinear Compton or Thomson scattering at a comparatively low centre-of-mass energy of the order of the electron mass, $\sqrt{s} \simeq 0.5$ MeV but rather high laser intensity, $I \simeq 100$ TW. Such an experiment is currently possible at the Forschungszentrum Dresden-Rossendorf (FZD), Germany, with the $40$ MeV linac ELBE~\cite{ELBE} and the $100$ TW laser DRACO \cite{Draco:2009}, so $\gamma = 80$ and a maximum value of $a_0 = 20$. The linac provides a well-defined electron beam with high brilliance and low emittance so that a detailed study of intensity effects on the radiation spectrum should become feasible. In addition, new technology developed for ELBE enables the delivery of ``bunches'' containing 1--10 electrons only. This could provide further new insights into the interaction of electrons with high intensity lasers.

Our paper is organised as follows. To make the presentation self-contained Section~\ref{sect.compton} briefly recapitulates the QED analysis of nonlinear Compton scattering. In Section~\ref{sect.thomson} this is compared with its classical limit (Thomson scattering). Section \ref{sect.radiation} briefly discusses how to obtain the spectrum from a classical radiation point of view. Effects caused by the finite temporal and spatial extent of the laser and electron pulses are discussed in Section~\ref{sect.pulse}. Section~\ref{sect.summary} contains our conclusions and summary. Some calculational details are relegated to the Appendices A and B.

\section{Nonlinear Compton Scattering}
\label{sect.compton}

\subsection{Kinematics}
\label{sect.compton.kinematics}

\begin{figure}[!ht]
\begin{center}
\includegraphics[scale=0.7]{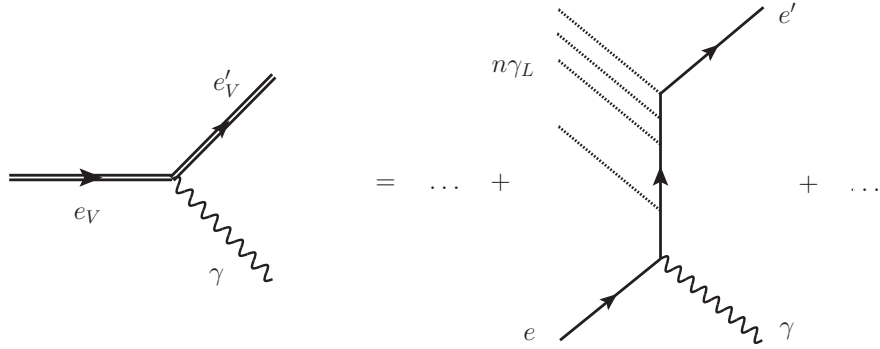}
\caption{Feynman diagram for nonlinear Compton scattering (left-hand side) in terms of dressed electron lines (subscript $V$). This diagram may be expanded in an infinite series of standard QED diagrams in which the $\ell$-th term corresponds to the absorption of $\ell$ laser photons $\gamma_L$ (right-hand side).} \label{fig.feynm.compton}
\end{center}
\end{figure}

Nonlinear Compton scattering, i.e.~the processes (\ref{reaction.nlcompton}) with $\ell > 1$, is treated in the text \cite{Berestetsky:1982aq} following \cite{Narozhnyi:1964} and has been reanalysed in \cite{Harvey:2009ry}. In this section we briefly collect the main results of the latter reference for later comparison, at the same time streamlining notation and normalisation conventions. Because of the large intensities involved one has to account for the possibility that many harmonic numbers $\ell$ contribute to the total cross section. In other words, nonlinear Compton scattering is nonperturbative in both the electromagnetic coupling $\alpha$ and laser intensity $a_0$ in that one has to sum over all individual processes (\ref{reaction.nlcompton}). In terms of Feynman diagrams this is depicted in Fig.~\ref{fig.feynm.compton}.

On the right-hand side we have displayed one term of the sum, the $\ell$-photon contribution, a tree level diagram with dotted lines representing laser photons, solid lines electrons, and wiggly lines the emitted photons. The nonperturbative sum is depicted on the left-hand side, the double lines denoting effective or dressed electrons corresponding to the Volkov solution of the Dirac equation in a plane wave \cite{Volkov:1935}. We are thus adopting a Furry picture \cite{Furry:1951zz} where the interaction with the classical laser background is shuffled into the ``free part'' of the Hamiltonian having the Volkov states as its `unperturbed' eigenstates. Schematically, the associated Volkov wave functions may be written as
\begin{equation} \label{VOLKOV}
    \Psi_{\ssvc{p}} = \exp \{i S[e; A]\} \, \Gamma[e; A] \, u_{\ssvc{p}} \; ,
\end{equation}
where $u_{\ssvc{p}}$ is a free Dirac spinor, $\Gamma[e; A]$ a field dependent combination of Dirac matrices and $S[e;A]$ the classical Hamilton-Jacobi action functional for a charge $e$ in a plane wave field $A$ \cite{Berestetsky:1982aq}. This suggests that the Volkov wavefunction is a WKB type solution of the Dirac equation. As expected from the Furry picture argument it contains all powers of the electromagnetic coupling $e$.

Throughout this section we assume the laser photons to be circularly polarised which, for a harmonic plane wave, corresponds to the electromagnetic 4-potential\footnote{Linear polarisation is obtained by simply setting one of the $\epsilon$'s to zero. The dimensionless intensities are then $a_0^2 = e^2 a^2/m^2$ and $a_0^2 = e^2 a^2/2 m^2$ for circular and linear polarisation, respectively.}
\begin{equation} \label{CIRC.P.W.}
  A^\mu = A^\mu (k \cdot x) = a (\epsilon_1^\mu \cos k\cdot x +  \epsilon_2^\mu \sin k\cdot x )\; ,
\end{equation}
where $k^2 = 0 = \epsilon_i \cdot k$ and $\epsilon_i \cdot \epsilon_j = - \delta_{i j}$. Assuming a head-on collision and summing over polarisation and spin states we thus expect the emission probabilities to be axially symmetric as there is no preferred direction other than the beam axis. Note that this is different for linear polarisation implying the azimuthal intensity patterns recorded in \cite{Chen:1998,Babzien:2006zz}.  On a more speculative level preferred directions along with azimuthal dependence may be induced by Lorentz violating physics such as noncommutative space-times \cite{Heinzl:2009zd}.

Suppressing electron spins the left-hand side of Fig.~\ref{fig.feynm.compton} may be analytically expressed as the S-matrix element
\begin{equation} \label{S-MATRIX}
  \langle \vcb{p}' ; \vcb{k}', \epsilon' \, | \, \mathbb{S} [A] \, |\vcb{p} \rangle  = -i e \int d^4x \, \overline{\Psi}_{\ssvc{p}'}(x) \
  \frac{e^{i k'\cdot \, x}}{\sqrt{2|\vcb{k}'|}}  \, \slashed{\epsilon}' \, \Psi_{\ssvc{p}}(x) \; ,
\end{equation}
where $\epsilon_\mu^\prime$ is the polarisation vector of the emitted photon, $k = (\omega, \vcb{k})$ and $k' = (\omega' , \vcb{k}') $ are the momenta of laser and emitted photons, respectively, and $p$ is the momentum of the incoming electrons before they actually enter the laser beam (assuming long but finite wave trains or adiabatic switching at infinite past and future \cite{Kibble:1965zz,Neville:1971uc}).

As already mentioned in the introduction the analogous diagram in QED (with ``naked" Dirac electrons) vanishes as one cannot satisfy $p + k = k'$ with all three particles being on-shell. This is different in the presence of an external field which can transfer additional 4-momentum. This statement seems obvious from the right-hand side of Fig.~\ref{fig.feynm.compton}. There is, however, an additional subtlety associated with the momentum assignment: in a plane wave field electrons acquire a quasi 4-momentum $q$ reflecting the relativistic quiver motion. This may be seen by calculating the proper-time averaged `Volkov current', $j^\mu_V \equiv \langle \bar{\Psi}_{\ssvc{p}} \gamma^\mu \Psi_{\ssvc{p}} \rangle = q^\mu/p^0$ as in \cite{Berestetsky:1982aq} or by writing the Hamilton-Jacobi action in (\ref{VOLKOV}) as $S[e;A] = q \cdot x + O((k \cdot x)^2)$. In either case the resulting quasi momentum is the intensity dependent quantity
\begin{equation}
  q \equiv p + \frac{a_0^2 \, m^2}{2k \cdot p} \, k,
\label{QUASI.MOMENTUM}
\end{equation}
and analogously for the outgoing electron momentum $q'$ (with $p$ replaced by $p'$). Working out the S-matrix element (\ref{S-MATRIX}) then yields a momentum conserving delta function with its support defined through
\begin{equation} \label{}
    P \equiv q + \ell k = q' + k' \; .
\end{equation}
The explicit dependence of quasi momenta on $a_0$ feeds through to the analogue of the Compton formula for the scattered frequency $\omega'$. This is compactly expressed in terms of dimensionless scalar products upon writing the relevant four-momenta as
\be \label{4VECTORS}
  k \equiv \omega n \; , \quad k' \equiv \omega' n' \; , \quad p \equiv m u = \gamma m (1, \vc{\beta}) \; .
\ee
Note that only the velocity $u$ transforms as a four-vector proper as only $m$ is a world scalar. It is useful to measure frequencies in units of the electron mass, $\omega \equiv \nu m$, $\omega' \equiv \nu' m$. Employing momentum conservation and the definitions (\ref{4VECTORS}) the (dimensionless) scattered frequency becomes
\be \label{NUPRIME.GENERAL}
  \nu_\ell^\prime = \frac{\ell \nu \, n \cdot u}{n' \cdot u + \displaystyle \left( \ell \nu + \frac{a_0^2}{2 n \cdot u} \right) n
  \cdot n'} \; .
\ee
For the time being we will focus on head-on collisions where $u = \gamma (1, -\beta \vcb{n})$ such that one can eliminate $n' \cdot u = n \cdot u - \beta\gamma \, n \cdot n'$ and (\ref{NUPRIME.GENERAL}) simplifies to
\begin{equation}\label{OMEGAPRIME}
    \nu'_\ell = \frac{\ell \nu n \cdot u}{n \cdot u + \kappa_\ell(a_0) \, n \cdot n'}  = \frac{\ell \nu}{1 + \kappa_\ell(a_0) \,  e^{-\zeta} \, (1 +
    \cos\theta)} \; .
\end{equation}
Here, $e^{-\zeta} = 1/ n \cdot u = \gamma (1 - \beta)$ is the usual Doppler factor, $\theta$ the scattering angle and
\begin{equation}\label{KAPPA.N}
    \kappa_\ell (a_0) \equiv  \ell \nu - \sinh \zeta + \frac{1}{2}a_0^2 \, e^{-\zeta} \equiv - \vcb{n} \cdot \vcb{P}/m
\end{equation}
is the projection of the total momentum $\vcb{P} = \vcb{q} + \ell \vcb{k}$ onto the optical axis, $\vcb{n} = \vcb{k}/\omega$, measured in units of $m$. The vanishing of the latter, $\vcb{n} \cdot \vcb{P} = - m \kappa_\ell = 0$, defines an intensity dependent centre-of-mass frame in which the scattered frequencies are precisely the harmonic multiples, $\nu_\ell^\prime = \ell \nu$ \cite{Harvey:2009ry}. For $\kappa_\ell < 0$ one is in the inverse Compton regime where the electrons transfer energy to the emitted photons, $\nu_\ell^\prime >  \ell \nu$, thus causing an overall blue-shift. This Doppler upshift, of course, is the physical basis for Compton generated X-rays.  The maximum scattered frequency, i.e.\ the Compton edge, occurs upon backscattering ($\theta=0$) and is given by
\begin{equation}\label{OMEGA.MAX}
    \nu_{\ell , \mathrm{max}}^\prime = \frac{\ell \nu e^{2\zeta}}{1 + 2\ell \nu e^\zeta + a_0^2} \; .
\end{equation}
The presence of $a_0$ in (\ref{OMEGA.MAX}) leads to a \textit{red-shift} of the $n=1$ Compton edge compared to linear Compton scattering ($a_0 \to 0$) which, for $\gamma \gg 1$, translates into the inequality
\be
  \nu_{1 , \mathrm{max}}^\prime (a_0) \simeq \frac{4 \gamma^2 \nu }{1 + 4 \gamma \nu + a_0^2} <  \frac{4 \gamma^2 \nu }{1 + 4 \gamma \nu} \simeq \nu_{1 , \mathrm{max}}^\prime (0) \; .
\ee
Hence, if one is primarily interested in up-shifting the laser frequency (say, for X-ray generation), the intensity $a_0$ should certainly not exceed unity. For large $a_0$ the Doppler upshift may even be completely compensated due to the ``stiffness'' of the laser beam whereupon one leaves the inverse scattering regime. As shown in \cite{Harvey:2009ry}, for $\gamma \gg 1$, this happens when $a_0$ exceeds a critical value of $2 \gamma$.

\subsection{Emission rates and cross sections}
\label{sect.compton.emission}

The $S$-matrix element (\ref{S-MATRIX}), represented by the Feynman diagram of Fig.~\ref{fig.feynm.compton} (left-hand side) is readily translated into an emission rate \cite{Berestetsky:1982aq,Harvey:2009ry}. The differential rate (per volume and time) for emitting a single photon of frequency $\omega' = m\nu'$ in the process (\ref{reaction.nlcompton}) is given by \cite{Narozhnyi:1964}
\begin{equation}
  \frac{d W_\ell}{d x} = \frac{e^2 m^2}{16 \pi} \, \frac{n_e}{q^0} \, a_0^2 \; \frac{\mathfrak{J}_\ell (z_\ell (x))}{(1+x)^2} \; ,
\label{eq.compton.emission_rate}
\end{equation}
where $n_e$ is the density of incoming electrons. We have chosen a normalisation somewhat different from \cite{Harvey:2009ry} such that $d W_\ell$ now has dimensions of $L^{-4}$ (i.e.\ particles per unit time and volume). The nontrivial part of the rate is encoded in the function ${\mathfrak J}_\ell$,
\begin{equation} \label{J}
  {\mathfrak J}_\ell(z_\ell) \equiv - \frac{4}{a_0^2} \, J^2_\ell(z_\ell) + \left(2 +
  \frac{x^2}{1 + x} \right) \Big[ J^2_{\ell-1}(z_\ell) + J^2_{\ell+1}(z_\ell) - 2J^2_\ell(z_\ell) \Big] \; .
\end{equation}
The $J_\ell$ are the usual Bessel functions of the first kind depending on the invariant argument
\begin{equation} \label{ZN}
  z_\ell (x) \equiv 2\ell \, \frac{a_0}{\sqrt{1 + a_0^2}} \, \sqrt{\frac{x}{y_\ell} \, \left( 1 -  \frac{x}{y_\ell} \right)}
\end{equation}
which is composed of two further invariants, namely
\begin{equation}\label{INVARIANTS}
  x \equiv \frac{k \cdot k'}{k \cdot p'} \; , \quad y_\ell \equiv \ell y_1 \equiv \frac{2\ell \, k \cdot p}{m_*^2} \; , \quad  (0 \le x \le y_\ell) \;
  .
\end{equation}
Note that $z_\ell = 0$ when $x$ acquires its minimal or maximal value. If we express the laser intensity in terms of (laser) photon density, $n_L$,
\begin{equation}\label{}
    a_0^2 = \frac{e^2 n_L}{m^2 \omega}  \; ,
\end{equation}
the differential rate (\ref{eq.compton.emission_rate}) may be written in a more symmetrical way,
\begin{equation}\label{W.ELL}
   \frac{d W_\ell}{d x} = r_e^2 \pi \,  \frac{n_L n_e}{k^0 q^0} \, m^2 \, \frac{\mathfrak{J}_\ell (z)}{(1+x)^2} \; ,
\end{equation}
$r_e = \alpha/m \simeq 3 \,\rm fm$ being the classical electron radius. Expressed in this form the rate is readily transformed into a cross section upon dividing by the (symmetric) flux factor \cite{Ivanov:2004fi},
\begin{equation}\label{FLUX.FACTOR}
    \jmath \equiv \frac{n_L n_e}{k^0 q^0} \, k \cdot q = \frac{n_L n_e}{k^0 q^0} \, m^2 \, \nu e^\zeta \; ,
\end{equation}
where the last identity holds for a head-on collision. We thus end up with the differential cross section
\be
  \frac{d\sigma_\ell}{d x} = r_e^2 \pi \, \frac{m^2}{k \cdot p} \, \frac{\mathfrak{J}_\ell (z)}{(1+x)^2} \; ,
\ee
which indeed has the correct dimensions of an area. Expanding $\mathfrak{J}_\ell (z)$ for small $a_0$ one recovers the Klein-Nishina cross section \cite{Narozhnyi:1964,Berestetsky:1982aq}\footnote{Note that \cite{Berestetsky:1982aq} has slightly different conventions: the densities $n_L$ and $n_e$ are set to unity, and Gaussian (rather than Heaviside-Lorentz) units are used which amounts to a reshuffling of factors of $4\pi$.}.

\begin{figure}[!ht]
 \includegraphics[scale=0.38,angle=270]{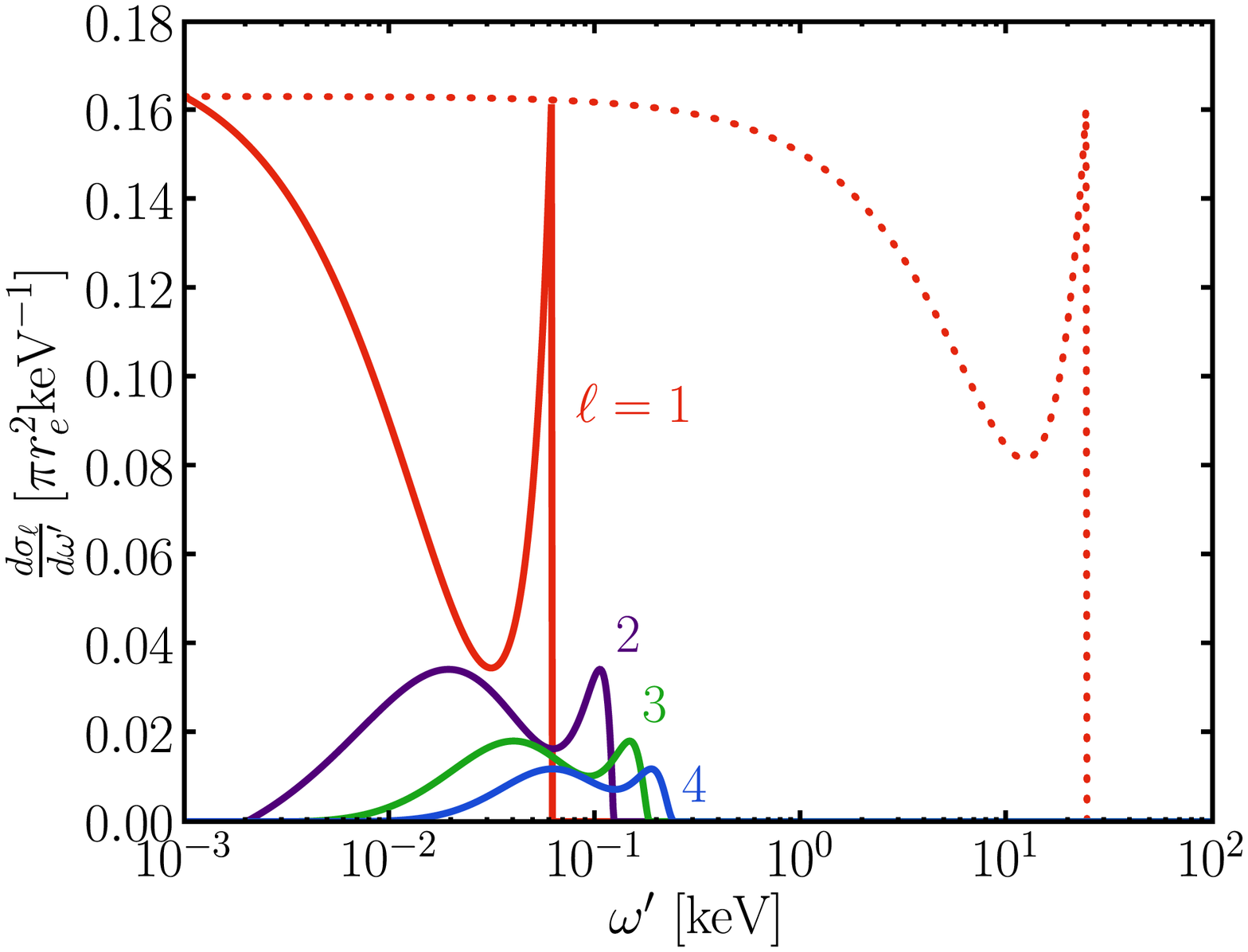}
 \includegraphics[scale=0.38,angle=270]{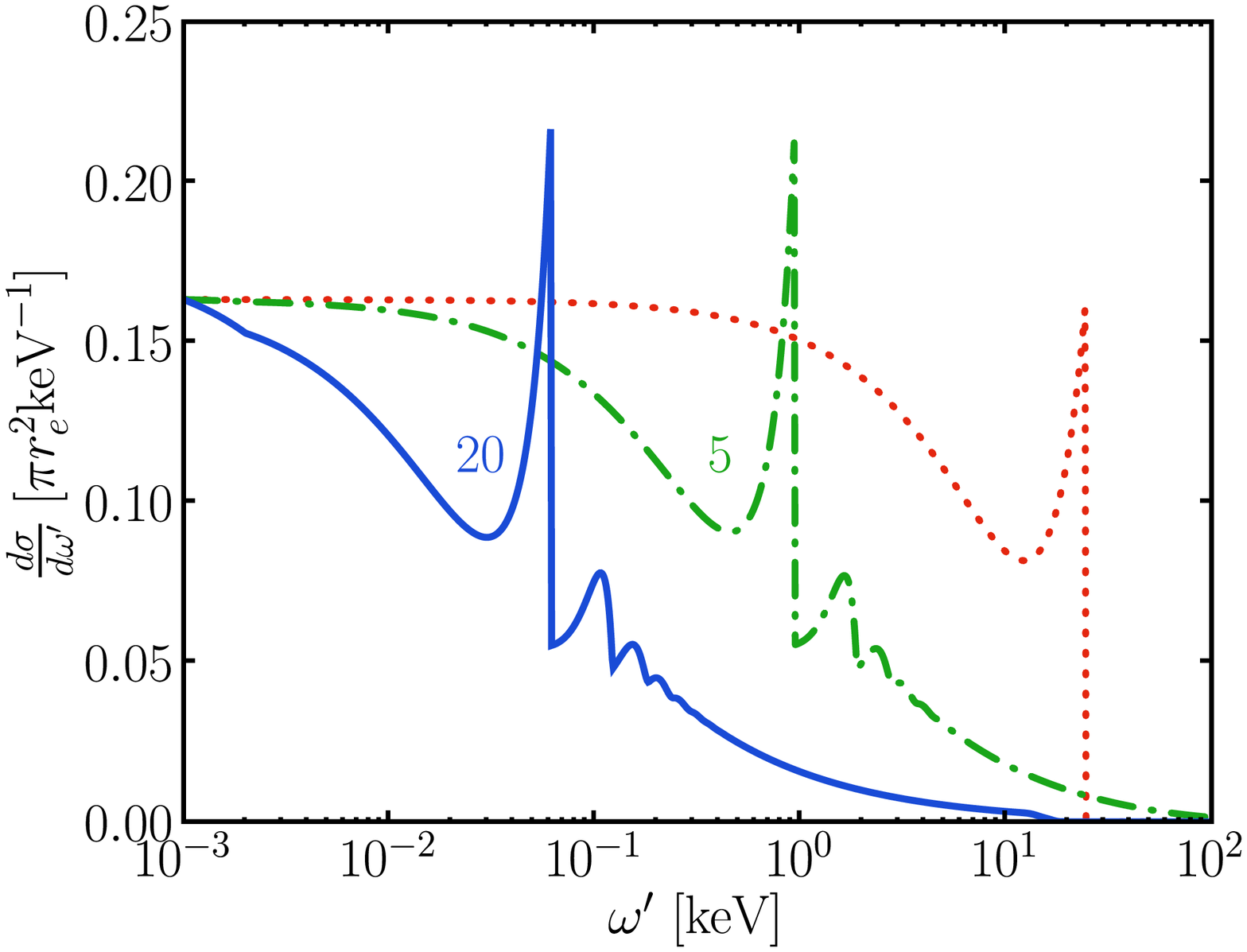}
 \caption{Left panel: Partial differential cross sections for the first few harmonics in nonlinear Compton back-scattering (head-on collision) for intensity $a_0 = 20$ (solid curve). Right panel: Harmonics summed up to $\ell = 500$ for intensities $a_0 = 20$ (solid curve) and $a_0 = 5$ (dash-dotted line). The spectrum for linear Compton scattering (Klein-Nishina formula) is displayed in both panels for comparison (dotted curve).}
 \label{fig.nlcompton_crossect}
\end{figure}

For what follows we will need rates and cross sections in the lab frame where we assume a head-on collision between laser and electron beams. The scattered frequency is then given by (\ref{OMEGAPRIME}) and the invariants (\ref{INVARIANTS}) become
\be
  x = \frac{\nu' (1 + \cos \theta)}{e^\zeta - \nu' (1 + \cos \theta)} \; , \quad y_1 = \frac{2\nu e^\zeta}{1 + a_0^2} \; .
\ee
This yields the following differential cross sections
\bea
  \frac{d \sigma_\ell}{d\omega'} &=& \frac{r_e^2 \pi}{\vcb{n}\cdot \vcb{P}} \, \frac{m^2}{k \cdot p} \, \mathfrak{J}_\ell (z_\ell) \; ,
  \label{D.SIGMA.FREQUENCY} \\
  \frac{d \sigma_\ell}{d\Omega}  &=& \frac{r_e^2}{2\ell} \, \left(\frac{\nu^\prime}{\nu e^{\zeta}}\right)^2 \, \mathfrak{J}_\ell (z_\ell) \; .
  \label{D.SIGMA.SOLID}
\eea
The individual harmonic cross sections (\ref{D.SIGMA.FREQUENCY}) are plotted in Fig.~\ref{fig.nlcompton_crossect} (left panel). One clearly sees that the contribution of each harmonic has its own frequency range given by $\ell \omega \le \omega_\ell^\prime \le \omega_{\ell, \mathrm{max}}^\prime$, cf.\ (\ref{OMEGA.MAX}), with the individual supports overlapping to some extent. The contributions of higher harmonics, $\ell > 1$, are getting more and more suppressed in amplitude. These features are sufficient to guarantee convergence of the  cross section summed over all harmonics \cite{Harvey:2009ry},
\begin{equation} \label{FULL.SIGMA}
  d\sigma \equiv \sum_{\ell = 1}^{\infty} d\sigma_\ell \; .
\end{equation}
The result of the summation (up to $n=500$) is shown in Fig.~\ref{fig.nlcompton_crossect} (right panel) for intensities $a_0 = 5$ and $a_0 = 20$. The figures basically coincide with those in \cite{Harvey:2009ry}, but now we are able to state the absolute normalisation on the vertical axis.

As already pointed out in \cite{Harvey:2009ry}, two main features can be seen to arise. First, in line with our discussion at the end of the previous subsection, the linear Compton 'edge' is red-shifted by a factor of $a_0^2 \simeq 400$ ($a_0^2 \simeq 25$) from about $4 \gamma^2 \omega \simeq 25$ keV down to $0.06$ keV ($1$ keV), i.e.\ from the hard to the soft X-ray regime when $a_0 = 20$. This is a rather drastic effect and it should be straightforward to verify experimentally \cite{Harvey:2009ry}. Second,  higher harmonics show up as additional peaks in the summed cross section with the peak heights decreasing rapidly with $\ell$.

\begin{figure}[!ht]
 \includegraphics[scale=0.45,angle=-90]{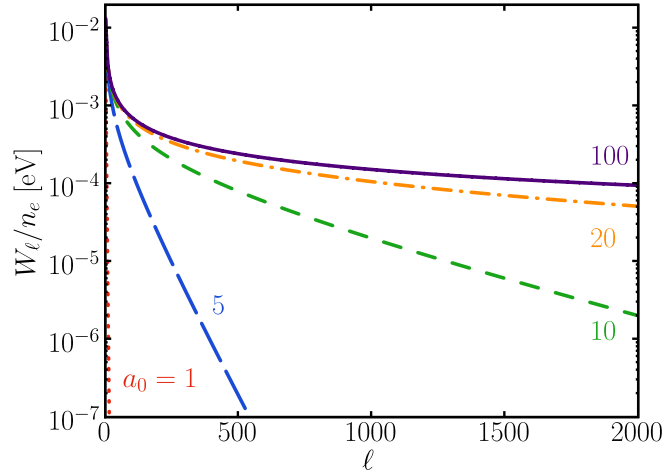}
 \caption{The integrated emission probability $W_\ell = \int d W_\ell$ from (\ref{W.ELL}) as a function of the harmonic number $\ell$ for various intensity parameters, $a_0=1, 5, 10, 20$ and $100$.} \label{fig.high.harmonics}
\end{figure}

We conclude this section by pointing out that the observation of high harmonics may become feasible, albeit implicitly,  using the \textit{total} emission rate. For sufficiently large values of $a_0$, very high harmonics, say even with $\ell > 1000$ have considerable integrated strength as shown in Fig.~\ref{fig.high.harmonics}. For instance, when $a_0 = 100$ and $\ell=2000$, the emission probability $W_\ell$ is only about two orders of magnitude below the value for the first harmonic, $\ell=1$. As Fig.~\ref{fig.high.harmonics} shows, summing such a large number of individual harmonics yields a rather smooth energy spectrum in total.

For a more detailed analysis of nonlinear Compton scattering, in particular its lab frame signatures, we refer the reader to \cite{Harvey:2009ry}. A complete discussion of polarisation effects has been given recently in \cite{Ivanov:2004fi}.

\section{Nonlinear Thomson Scattering}
\label{sect.thomson}

In this section we compare the results for nonlinear Compton scattering from the previous Section with the results for nonlinear Thomson scattering considered, for instance, in \cite{Sarachik:1970ap,Esarey:1993zz}. As stated in \cite{Nikishov:1963,Narozhnyi:1964}, and further analysed in \cite{Harvey:2009ry}, the Thomson limit is obtained when the invariant $y_\ell$ defined in (\ref{INVARIANTS}) becomes small. For head-on kinematics this is the statement
\be \label{RECOIL}
  y_\ell (a_0) = \frac{2\ell \, \nu e^\zeta}{1 + a_0^2} \ll 1 \; .
\ee
Physically, the quantity $y_\ell$ represents the maximum (normalised) recoil of the electron during the scattering process, Hence, the Thomson limit amounts to neglecting the momentum transfer from the laser field to the electron. From (\ref{RECOIL}) this is feasible if either $e^\zeta \simeq 2 \gamma$ and $\ell$ are sufficiently small or for large intensity, $a_0^2$, i.e.\ large photon density. The latter is, of course, consistent with approaching the classical limit. 

A closely related quantity measuring quantum nonlinear effects has also been introduced in \cite{Nikishov:1963,Narozhnyi:1964}, namely
\be \label{CHI}
  \chi \equiv \frac{e \sqrt{F^{\mu\nu}p_\nu}}{m^3} = \frac{k \cdot p}{m^2} \, a_0 = \frac{1}{2} \, y_1(0) \, a_0 \; . 
\ee
This has frequently being used since then, but as the last identity in (\ref{CHI}) shows it is basically equivalent to (\ref{RECOIL}). Following the example of \cite{Narozhnyi:1964} we prefer to work with $y_\ell$ in what follows\footnote{The second equality in (\ref{CHI}) follows from the gauge invariant definition of $a_0$, see e.g.\ \cite{Heinzl:2008rh}}.

Finally, it is worth noting that $y_\ell$ is related to the usual Mandelstam variable
\be
  s = (q + \ell k)^2 = (q' + k')^2  = m_*^2 (1 + y_\ell) \; ,
\ee
representing the total centre-of-mass energy squared. Hence, in the Thomson limit, one neglects the photon contributions to this energy, so that $m_*$ is the dominant energy scale, $s \simeq m_*^2$. In this sense, the Thomson limit is a low-energy limit. In contrast, the SLAC E-144 experiment \cite{Bamber:1999zt} ($\gamma \simeq 10^5$, $a_0 \simeq 0.6$, hence $y_1 \simeq 1$) has probed the genuine Compton (or quantum) regime.

In what follows we want to find explicit relations between the general Compton expressions and their classical (Thomson) limit. To this end we try to separate off the quantum corrections from the purely classical results. We begin with the momentum projection (\ref{KAPPA.N}) which may be rewritten as
\be
  \kappa_\ell = \kappa_0 + \frac{1}{2} y_\ell (1 + a_0^2) e^{-\zeta} \; ,
\ee
where $\kappa_0$ is obtained by setting $\ell = 0$. Replacing $\kappa_\ell \to \kappa_0$ in the scattered frequency (\ref{OMEGAPRIME}) for head-on collisions straightforwardly yields the (nonlinear) Thomson limit,
\be \label{NUPRIME.TH}
  \nu_{\ell,\mathrm{Th}}^\prime = \frac{\ell \nu \, n \cdot u}{n \cdot u + \kappa_0 \, n \cdot n'} \equiv  \frac{\ell \nu}{1 + \kappa_0 e^{-\zeta} (1 + \cos \theta)} \equiv \ell \nu_1' \; .
\ee
This suggests that the general formula for arbitrary collision geometry is obtained by setting $\ell = 0$ in the denominator of (\ref{NUPRIME.GENERAL}) such that the scattered frequencies
\be \label{NUPRIME.CLASSICAL}
  \nu_{\ell, \mathrm{Th}}^\prime = \ell \nu_1' \equiv \frac{\ell \nu \, n \cdot u}{n' \cdot u + \displaystyle \frac{a_0^2}{2 n \cdot u}  \, n \cdot n'}  \;
\ee
are indeed integer multiples of a fundamental frequency $\nu_1'$. At this point it is instructive to compare with the low intensity (``linear'') limit (or Thomson limit proper) where $\ell = 1$, $a_0 \to 0$ and (\ref{NUPRIME.CLASSICAL}) condenses to
\be \label{NUPRIME.CLASSICAL.LINEAR}
  \nu' = \frac{n \cdot u}{n' \cdot u} \, \nu \; .
\ee
This is the Doppler shift in disguise upon noting that for a head-on collision and backscattering $n \cdot u = \gamma (1 + \beta) = e^\zeta$ and $n' \cdot u = \gamma (1 - \beta) = e^{-\zeta}$.

\begin{figure}[!ht]
 \includegraphics[scale=0.38,angle=-90]{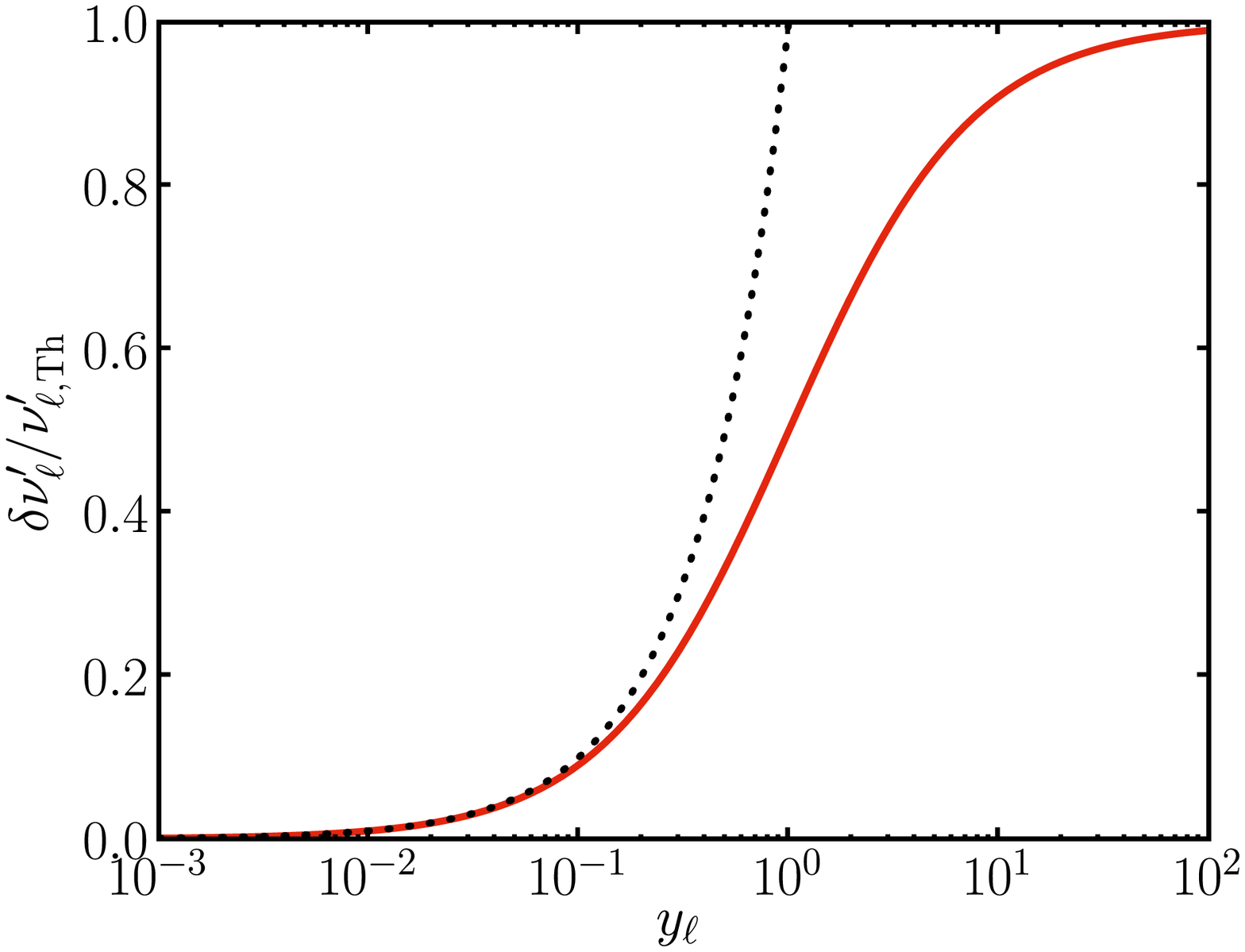}
 \includegraphics[scale=0.38,angle=-90]{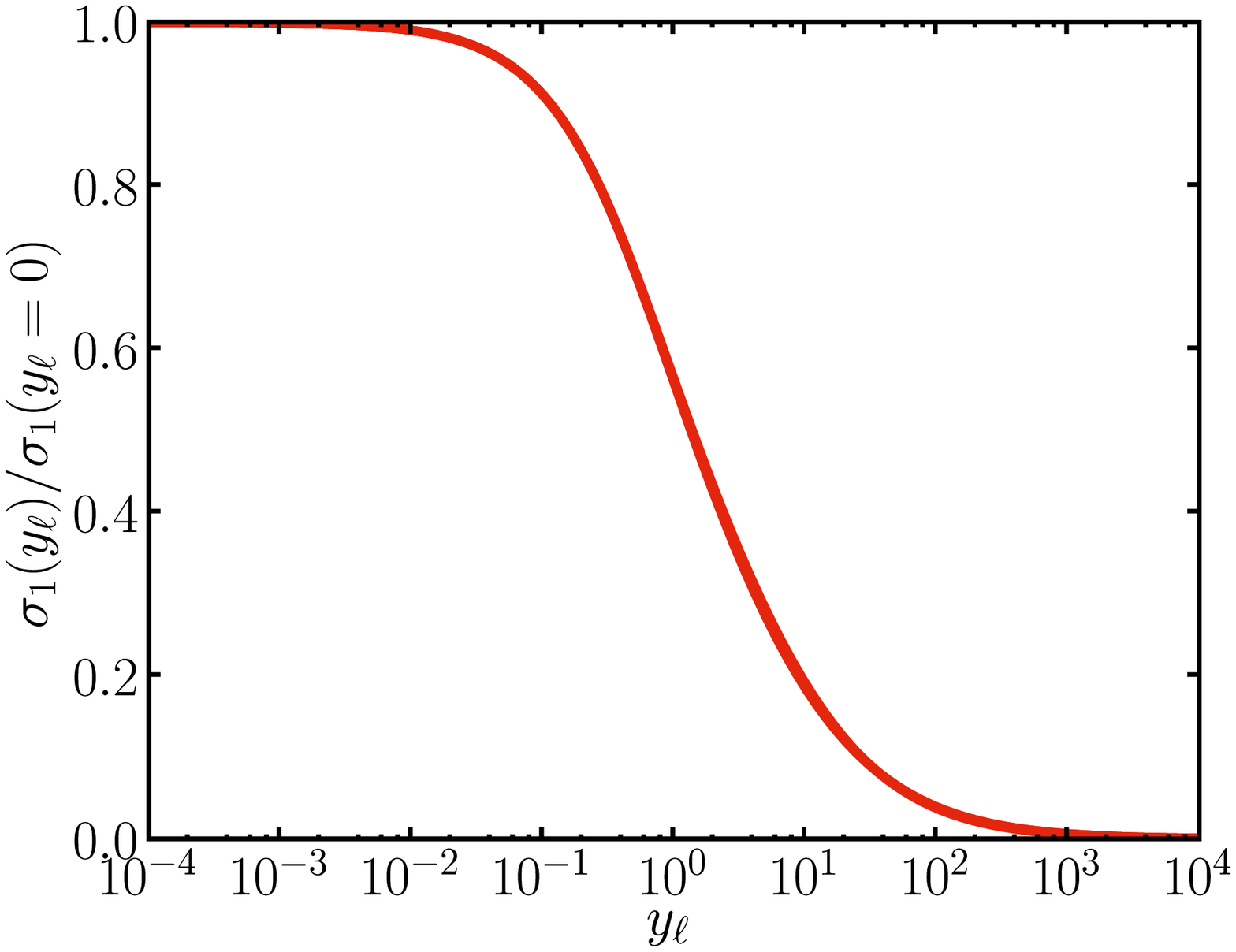}

\caption{Left panel: Comparison of classical (Thomson) and QED (Compton) results for the scattered frequency in the backscattering direction, $\theta=0$, with $\delta \nu^\prime_\ell \equiv \nu^\prime_{\ell, \mathrm{Th}} - \nu^\prime_\ell$ plotted as a function of $y_\ell$. Solid line: $x_\ell/(1 + x_\ell)$, cf.~(\ref{NUPRIME.C.TH}); dotted line: Thomson limit (\ref{NUPRIME.TH}). Right Panel: Normalised integrated partial cross section $\sigma_\ell$ as a function of the electron recoil $y_\ell$ normalised to Thomson limit, cf.~(\ref{CT.RATIO}).
Choosing $\ell = 1$ the curves for $a_0 \to 0$ and $a_0=5$ up to $20$  are on top of each other.}
\label{fig.classical_vs_qed}
\end{figure}

Expressing the invariant $x$ from (\ref{INVARIANTS}) in terms of the scattering angle $\theta$ it becomes explicitly $\ell$-dependent \cite{Harvey:2009ry}, $x \equiv x_\ell = \ell x_1$, with
\be
   x_1 = \frac{\nu e^{-\zeta} (1 + \cos \theta)}{1 + \kappa_0 e^{-\zeta} (1 + \cos \theta)} \; .
\ee
Comparing with (\ref{NUPRIME.TH}) one finds the relation
\be \label{NUPRIME.C.TH}
  \nu_\ell' = \frac{\nu_{\ell,\mathrm{Th}}^\prime}{1 + x_\ell} \; .
\ee
As $x_\ell$ is bounded by $y_\ell$ the Thomson limit implies $x_\ell \to 0$ so that Compton and Thomson expressions should generically differ by terms of order $x_\ell$ as in (\ref{NUPRIME.C.TH}). This is consistent with the findings in \cite{Harvey:2009ry} and confirmed by the numerical comparison of Fig.~\ref{fig.classical_vs_qed}, left panel.

Let us move on to the emission rates. Again, we try to isolate all terms dependent on $x_\ell$. Following \cite{Harvey:2009ry} we define the ratio $r = x_1/y_1$ with $0 \le r \le 1$ and rewrite $z_\ell$ from (\ref{ZN}) as
\be
  z_\ell = \ell z_1 = 2\ell \, \frac{a_0}{\sqrt{1 + a_0^2}} \, \sqrt{r(1-r)} \; , \quad (0 \le z_1 < 1) \; .
\ee
It is important to take the Thomson limits, $x_1 \to 0$ and $y_1 \to 0$, in such a way that the ratio $r$ stays \textit{fixed} as a result of which $z_\ell$ remains unchanged. We may therefore decompose the spectral function (\ref{J}) into a classical (Thomson) part and an $x_\ell$ dependent correction,
\be
  \mathfrak{J}_\ell (z_\ell) = \mathfrak{K}_\ell (z_\ell) + \frac{x_\ell^2}{1 + x_\ell} \mathfrak{L}_\ell (z_\ell) \; ,
\ee
where (suppressing the overall argument $z_\ell$)
\be
  \mathfrak{L}_\ell \equiv J_{\ell-1}^2 + J_{\ell+1}^2 - 2 J_\ell^2 \; , \quad \mbox{and} \quad \mathfrak{K}_\ell \equiv -4 J_\ell^2/a_0^2 + 2
  \mathfrak{L}_\ell \; .
\ee
This yields the Thomson limit of the cross section (\ref{D.SIGMA.SOLID}),
\be
  \left( \frac{d \sigma_\ell}{d\Omega}\right)_{\mathrm{Th}} = \frac{r_e^2}{2\ell} \, \left(\frac{\nu^{\prime}}{\nu e^{\zeta}}\right)^2 \,
  \mathfrak{K}_\ell (z_\ell) \; ,
\ee
where only the classical part, $\mathfrak{K}_\ell$, of $\mathfrak{J}_\ell$ contributes. Factoring $\mathfrak{K}_\ell$ out from (\ref{D.SIGMA.SOLID}) we may explicitly calculate the Compton to Thomson ratio,
\be \label{CT.RATIO}
  \frac{d \sigma_\ell}{d\Omega} \bigg/ \left( \frac{d\sigma_\ell}{d\Omega}\right)_{\mathrm{Th}} = \frac{1}{(1 + x_\ell)^2} + \frac{x_\ell^2}{2(1 +
  x_\ell)^3} \left( 1 + \frac{4}{a_0^2}\frac{J_\ell^2}{\mathfrak{K}_\ell} \right) = 1 + O(x_\ell) \; .
\ee
Again, Compton and Thomson results differ by terms of order $x_\ell$. A numerical comparison is displayed in Fig.~\ref{fig.classical_vs_qed}, right panel.

For our set of parameters one finds $y_\ell \simeq 2 \times 10^{-6} \ell$ which is very small unless one considers extremely high harmonics ($\ell > 10^6$). Thus, according to (\ref{CT.RATIO}), Compton and Thomson results differ by approximately $10^{-5}$ for low harmonics. This suggests, for these parameters, that a purely classical calculation will yield a very good approximation for the Compton scattering results. This will be further corroborated in the following Section.

\section{Classical Radiation}
\label{sect.radiation}

Having considered the classical limit of Compton scattering it is worthwhile comparing with a purely classical calculation. This will later be used to investigate various finite size effects on the spectrum. We are particularly interested in a fully covariant description of the radiation emitted as bremsstrahlung by a charge moving in a laser field such as given by (\ref{CIRC.P.W.}). It is most convenient (and, as an additional bonus, also maintains explicit gauge invariance) to start from  the field strength tensor \cite{Heinzl:2008rh}
\be \label{FMUNU}
  F^{\mu\nu} = F_i (k \cdot x) \, f_i^{\mu\nu} \; , \quad f_i^{\mu\nu} = n^\mu \epsilon_i^\nu - n^\nu \epsilon_i^\mu \; , \quad i = 1,2 \; ,
\ee
with $n^\mu = k^\mu/\omega$ as before and $\epsilon_i^\mu$ the two transverse polarisation vectors introduced for the circularly polarised gauge field (\ref{CIRC.P.W.}) which corresponds to the choice\footnote{To obtain linear polarisation just set one of the $F$'s to zero.}
\be \label{F12}
  F_1 (k \cdot x) =  -a \omega \, \sin (k \cdot x) \; , \quad F_2 (k \cdot x) = a \omega \, \cos (k \cdot x) \; ,
\ee
such that $F \equiv a \omega$ represents the magnitude of electric and magnetic field.

The Lorentz equation of motion for a charge in a plane wave has first (and quite elegantly) been solved by Taub in 1948 \cite{Taub:1948} (for other early treatments see \cite{Sengupta:1949,Sarachik:1970ap,Meyer:1971pj}). A modern covariant analysis has recently appeared \cite{Heinzl:2008rh} on which the following remarks are based. For the following it is useful to absorb charge $e$ and mass $m$ into the field variables i.e.\ to replace $(e/m) F^{\mu\nu} \to F^{\mu\nu}$ and $(e/m) A^\mu \to A^\mu$ such that the rescaled $A^\mu$ is dimensionless\footnote{The intensity parameter hence becomes $a_0^2 \equiv - \langle A^2 \rangle = a^2$ for circular and $a_0^2 = a^2/2$ for linear polarisation, respectively.}. The Lorentz equation of motion then reads
\be \label{LORENTZ}
  \dot{p}^\mu = F^{\mu}_{\;\nu} (x(\tau)) \, p^\nu \; ,
\ee
and is, in general, a nonlinear differential equation as the field strength, $F^{\mu\nu}$, depends on the trajectory $x(\tau)$ to be solved for. Thus, normally, one cannot expect to find an analytic solution. For a plane wave, however, where $F^{\mu\nu} = F^{\mu\nu} (k \cdot x)$ there is a sufficient number of conserved quantities such that the system becomes integrable. Most important is the constancy of longitudinal momentum in time, $k \cdot p \equiv m \Omega = const$, which defines a frequency $\Omega = \omega n \cdot u $ such that $k \cdot x$ can be traded for proper time via
\be \label{CAP.OMEGA}
  k \cdot x = \Omega \tau = \omega \, n \cdot u \, \tau = \omega \, n \cdot u_0 \, \tau \; ,
\ee
adopting the initial conditions $k \cdot x(0) = 0$ and $u(0) = u_0$. Integrability is now obvious as $F^{\mu\nu} = F^{\mu\nu} (\Omega \tau)$ which turns (\ref{LORENTZ}) into a linear equation. All amplitude functions become functions solely of proper time. Integration of (\ref{LORENTZ}) is now straightforwardly done via exponentiation, $u = \exp (\int d\tau \, F) u_0$. The null field (\ref{FMUNU}) is nilpotent of order two, i.e.\ all powers higher than two vanish \cite{Heinzl:2008rh}. Thus the exponential series gets truncated after the term quadratic in field strength leaving us with
\be \label{UMU}
  u^\mu (\tau) = u_0^\mu - A^\mu  + \frac{2 A \cdot u_0 - A^2}{2 n \cdot u_0} \, n^\mu \; , \quad A^\mu = A^\mu (\tau) \; .
\ee
Note that we have assumed the initial condition $A^\mu (0) = 0$, otherwise one has to replace $A^\mu \to A^\mu - A^\mu (0)$. It is reassuring to directly check the conservation law $n \cdot u = n \cdot u_0$ which holds as $n \cdot A = 0$ (Lorenz or light-cone gauge) and $n^2 = 0$.

Another $\tau$ integration of (\ref{UMU}) then yields the orbit
\be \label{XMU}
  x^\mu (\tau) = x^\mu (0) + \tau u_0^\mu  - \int_0^\tau d\tau' A^\mu + \frac{1}{2 n \cdot u_0} \int_0^\tau d\tau' \, (2 A \cdot u_0 - A^2) \, n^\mu
  \; .
\ee
The first two terms are obviously initial conditions, followed by a transverse part along $A^\mu$ and a longitudinal contribution proportional to $n^\mu = k^\mu/\omega$.

The calculation of the classical radiation spectrum may be found in most texts on electrodynamics (see e.g. \cite{Jackson:1999,Landau:1987}) though not necessarily in a covariant manner. One has to determine the 4-momentum $P^\mu$ of the radiation field using Poynting's theorem and the retarded potentials, $A_{\mathrm{ret}}^\mu = \Box^{-1} j^\mu |_{\mathrm{ret}}$. A particularly compact covariant expression is given by the Fourier integral \cite{Mitter:1998}
\be \label{P.RAD}
  P^\mu = - \int \frac{d^4 k}{(2 \pi)^3} \, \mbox{sgn}(k^0) \, \delta (k^2) \, k^\mu \, j(k) \cdot j(-k) \; ,
\ee
where $j^{\mu} (k)$ is the Fourier transform of the electron current,
\be \label{J.MU.X}
  j^\mu (x) = e \int d X^\mu \, \delta^4 (x - X(\tau)) = e \int d\tau \, u^\mu (\tau) \delta^4 (x - X(\tau)) \; , \quad u^\mu \equiv d X^\mu/d\tau \; .
\ee
Accordingly, (\ref{P.RAD}) expresses the radiation 4-momentum in terms of the electron trajectory, $X = X(\tau)$, encoded in the current (\ref{J.MU.X}). The zero component of (\ref{P.RAD}), of course, yields the radiated energy which, after performing the $k^0$ integration, we write as
\be \label{P0}
  P^0 \equiv \int d\omega' d\Omega \,  \omega' \, \frac{d^2 N_\gamma (k')}{d\omega' d\Omega}  \; , \quad k' = \omega' (1, \vcb{n}') \; ,
\ee
where we have reinstated primes for scattered momentum components. The integrand is the spectral density describing the number of photons radiated per unit frequency per unit solid angle,
\be \label{RHO.DEF}
  \rho(\omega', \mathbf n') \equiv \frac{d^2 N_\gamma}{d\omega' d\Omega} = - \frac{\omega'}{16 \pi^3} \, j(k') \cdot j^* (k') \; .
\ee
Thus, to determine the spectral density all we need to know is the Fourier transform $j^\mu (k')$ of the current (\ref{J.MU.X}) which depends both on orbit position and velocity, $X^\mu$ and $u^\mu$, respectively. Employing the continuity equation, $k' \cdot j(k') = 0$, one may eliminate $j^0$ such that (\ref{RHO.DEF}) turns into
\be \label{RHO.3D}
  \rho(\omega', \vc{n}') = \frac{\omega'}{16 \pi^3} \, |\vcb{n}' \times \vcb{j}(k')|^2 \ge 0 \; .
\ee
In summary, a determination of the classical radiation spectrum in an external field amounts to solving the Lorentz force equation in this field for the trajectory, and hence, to find the current $j^\mu (x)$ of (\ref{J.MU.X}).  Its Fourier transform, compactly written as
\be \label{J.MU.K}
  j^\mu (k') = e \int d\tau \, u^\mu(\tau) e^{-i k' \cdot x(\tau)} \; ,
\ee
then yields the spectrum via the radiation formula (\ref{RHO.DEF}). The Fourier integral (\ref{J.MU.K})
may be (partly) evaluated using a trick of Schwinger's \cite{Schwinger:1949ym}, assuming periodic $\tau$ dependence of the gauge field as in (\ref{CIRC.P.W.}) and (\ref{F12}). In this case the four-velocity is periodic as well,
\be
  u^\mu (\tau + \ell' T) = u^\mu (\tau) \; , \quad T \equiv 2\pi/\Omega \; , \quad \ell' \; \mathrm{integer} \; ,
\ee
but for the orbit one finds
\be \label{ORBIT.PERIODIC}
  x^\mu (\tau + \ell' T) =  \ell' T w^\mu + x^\mu (\tau) \; ,
\ee
where $w^\mu$ is the velocity averaged over one period of proper time and hence proportional to the quasi-momentum (\ref{QUASI.MOMENTUM}), $q^\mu = m w^\mu$. Note that the dependence on $\ell'$ has been separated off in the first term of (\ref{ORBIT.PERIODIC}). For (almost) periodic functions $f(\tau)$ it makes sense to decompose the integral over all $\tau$ into a sum of integrals over all periods, i.e.
\be
  \int d\tau \, f(\tau) =  \sum_{\ell' = -\infty}^\infty \, \int_0^T d\tau \, f(\tau + \ell' T) \; .
\ee
Applying this to the current (\ref{J.MU.K}) and, once again, separating off the $\ell'$ dependent pieces one finds the expression
\be \label{J.MU.K.SUM}
  j^\mu (k') = e \sum_{\ell'} \exp \left\{ 2\pi i \ell' \, \frac{\omega'}{\Omega} \left( n' \cdot u_0 + \frac{a_0^2}{2 n \cdot u_0} n \cdot n' \right) \right\} j_T^\mu (\omega', n') \; ,
\ee
where we have defined the integral over a single period $T$,
\be \label{J.MU.K.SINGLE.PERIOD}
  j_T^\mu (\omega', n') \equiv \int_0^T d\tau \, u^\mu (\tau) \, \exp \left\{ -i \, \omega'\, n' \cdot x_0 -i \, \omega' \int_0^\tau d\tau' \, n' \cdot u(\tau') \right\} \; ,
\ee
which is independent of $\ell'$. Thus, in (\ref{J.MU.K.SUM}) all dependence on $\ell'$ has again been factored off such that the sum over all periods can be evaluated using Poisson resummation, $\sum_{\ell'} \exp(2\pi i \ell' h) = \sum_\ell \delta(\ell - h)$. As a result, we obtain a ``delta comb'' for the current,
\be \label{J.MU.K.RESULT}
  j^\mu (k') = \omega_1' \sum_{\ell > 0} \delta (\omega' - \ell \omega_1') \; j_T^\mu (\ell \omega_1', \vcb{n}') \; ,
\ee
where the multiples of $\omega_1' = m \nu_1'$, derived from the exponent in (\ref{J.MU.K.SINGLE.PERIOD}), define the harmonic frequencies $\nu_\ell^\prime = \ell \nu_1'$ which \textit{precisely} coincide with the Thomson limit (\ref{NUPRIME.CLASSICAL}) upon identifying $u_0$ with the asymptotic velocity $u=p/m$ of the scattering process\footnote{For head-on collisions we recover (\ref{NUPRIME.TH}).}. Mod-squaring our answer we thus conclude with Schwinger \cite{Schwinger:1949ym} that periodic motion (induced by periodic fields of infinite spatio-temporal extension) leads to a line spectrum of radiation into harmonics labelled by index $\ell$.

What can we expect to happen when we restrict the fields to have finite temporal duration, i.e.\ for pulses? In general, an analytic treatment will be difficult, but there is one particular case which is reasonably straightforward and nevertheless yields the basic physics involved. This is the case of a rectangular pulse (see also \cite{Krafft:2004}), where we just cut off the $\tau$ integral in (\ref{J.MU.K}) such that $-\tau_0/2 \le \tau \le \tau_0/2$. Accordingly, the sum over $\ell'$ in (\ref{J.MU.K.SUM}) only extends from  $-N$ to $N$, where we assume that our pulse contains $2N$ periods, $\tau_0/T = 2N$. The finite sum can nevertheless be evaluated with the result that (\ref{J.MU.K.RESULT}) gets replaced by
\be \label{JN.FINITE.N}
  j_N^\mu (k') = \frac{\sin(2N+1) \pi \displaystyle \frac{\omega'}{\omega_1'}  }{\sin\pi \displaystyle \frac{\omega'}{\omega_1'} } \, j_T^\mu (\omega', \vcb{n}') \; .
\ee
The prefactor composed of the ratio of two sines is nothing but the diffraction pattern obtained when light passes $2N+1$ slits. It has the usual maxima at $\omega'/\omega_1' = \ell$ integer but also $N-1$ \textit{additional peaks} of lower amplitude between adjacent integers. Thus, in our language, each harmonic is accompanied by a substructure of $N-1 = \tau_0/2T - 1$ secondary peaks. Furthermore, there is of course line broadening, as the (formal) zero-width limit of the delta comb is only achieved for infinite $N$, hence infinite temporal extent.

These findings should qualitatively also hold for more realistic (smooth) pulse shapes such as Gaussians or power laws. We will explore this in more detail in the next section.

Before we come to that we conclude the general discussion with a remark on the relevance of radiation damping. In principle, the Lorentz equation (\ref{LORENTZ}) is only valid approximately as it neglects the back-reaction of the radiation field on the electron orbit (see \cite{Gralla:2009md} for an illuminating recent discussion). To incorporate the latter the Lorentz equation is superseded by the Landau-Lifshitz one \cite{Landau:1987} which was recently utilised in the context of high-intensity lasers \cite{DiPiazza:2009pk}. For our parameters, the radiation reaction parameter defined there becomes
\be
  R = \frac{4}{3} \alpha \gamma \nu a_0^2 \simeq 10^{-3} \; .
\ee
We have checked that for our electron beam and laser pulse specifications the longitudinal motion is barely altered. In particular, radiation damping modifies the change in energy after collision by just 0.58 \%. Accordingly, the backscattered photon spectrum will change at the sub-percent level at most, and we can safely neglect back-reaction effects. For more detailed discussions of radiation damping in a laser context the reader is referred to \cite{Sarachik:1970ap,Hartemann:1996zza,DiPiazza:2009pk,Sokolov:2009}.

\section{Beam shape effects}
\label{sect.pulse}

It has become clear that treating the laser beam as a plane wave of infinite spatial and temporal extent according to (\ref{CIRC.P.W.}) and (\ref{F12}) is an idealisation, the validity of which has to be checked. An infinite plane wave should be a fair approximation if the electrons only probe the central region of the laser focus and if the reaction time is small compared to pulse duration $\sigma$. To assess the feasibility of these assumptions and the size of possible modifications we will study finite size effects by first considering pulsed plane waves and, in a second step, allowing for a transverse intensity profile in addition.

\subsection{Finite temporal duration}
\label{sect.thomson.pulse}

A pulsed plane wave is obtained upon multiplying the field strength $F^{\mu\nu}$ from (\ref{FMUNU}) corresponding to the infinite wave (\ref{CIRC.P.W.}) with an envelope factor of width $\sigma$ such that (\ref{F12}) gets replaced by
\be \label{CIRC.PULSE}
  F_1 (k \cdot x) = -g_\sigma (k \cdot x) \, F \, \sin (k \cdot x) \; , \quad F_2 (k \cdot x) = g_\sigma (k \cdot x) \, F \, \cos (k
  \cdot x) \; ,
\ee
with some suitable envelope factor $g_\sigma$ and $F= a\omega$. Note that this does not spoil the fact that the associated gauge potential $A^\mu$ (in radiation gauge) still represents a plane wave solution of Maxwell's equations, as $k^2 = \omega^2 n^2 = 0 $ implies $\Box A^\mu (k \cdot x) = 0$. By writing the pulse width as $\sigma \equiv \Omega \tau_0$ with a pulse duration $\tau_0$ in proper time, we have
\be
  g_\sigma(k \cdot x) = g(\tau/\tau_0) \; .
\ee
Orbit velocity and trajectory are still given by (\ref{UMU}) and (\ref{XMU}) if the pulse factors are included in the definition of the gauge potential $A^\mu (\tau)$. Hence, a useful analytic expression for the orbit can be obtained whenever the $\tau$ integrals over $A$ and $A^2$ can be evaluated which, of course, crucially depends on the pulse shape function $g$. It has been suggested \cite{McDonald:1997} to use the ``solitonic'' pulse
\be \label{SECH}
  g (\tau) = \sech (\tau/\tau_0) \; ,
\ee
which, being the generating function of Euler numbers, does not lend itself to straightforward integrations. A particularly simple case, however, is obtained by considering crossed fields with $F^{\mu\nu}$ constant, hence $A^\mu$ linear in $\tau$,
\be
  F^{\mu\nu} = F \, f^{\mu\nu} \; , \quad A^\mu = a \Omega \tau \, \epsilon^\mu \; .
\ee
Superimposing a power-law pulse shape, the amplitudes become time-dependent, e.g.
\be
  F^{\mu\nu} (\tau) = \frac{F}{(1 + \tau^2/\tau_0^2)^{3/2}} \, f^{\mu\nu} \; , \quad  A^\mu (\tau)  = \frac{a\Omega \tau}{(1 + \tau^2/\tau_0^2)^{1/2}} \, \epsilon^\mu \; ,
\ee
such that the orbit coefficient functions appearing in (\ref{XMU}) become
\bea
  \int_0^\tau d\tau' \, A^\mu (\tau') &=& a \Omega \tau_0^2 \epsilon^\mu\, \left( \sqrt{1 + \tau^2/\tau_0^2} - 1 \right) \; , \label{b1}\\
  \int_0^\tau d\tau' \, A^2 (\tau') &=& -a^2 \Omega^2 \tau_0^3 \left( \frac{\tau}{\tau_0} - \arctan \frac{\tau}{\tau_0} \right) \label{b}\; .
\eea
As a result we note that within the pulse ($\tau \ll \tau_0$) the coefficients (\ref{b1}) and (\ref{b}) are quadratic and cubic in $\tau$, respectively, hence behave as for infinite crossed fields. Outside the pulse, i.e. when it has passed by ($\tau \gg \tau_0$), the orbits become modified and approach free motion linear in $\tau$.

As we have seen in the previous section, for any particular choice of amplitudes one has to insert the velocity (\ref{UMU}) and the orbit (\ref{XMU}) into (\ref{J.MU.K}) and evaluate the radiation formula (\ref{RHO.DEF}) to analyse the influence of the finite temporal extent encoded in the pulse shape $g(\tau)$. For a rectangular pulse we saw secondary peaks appearing in a way reminiscent of a diffraction pattern.

For smooth pulse shapes $g$ the situation is slightly different. Adopting backscattering kinematics for simplicity and the circularly polarised pulse (\ref{CIRC.PULSE}), the electron current (\ref{J.MU.K}) becomes
\begin{eqnarray} \label{J.K.SMOOTH}
  \mathbf j(\omega') &=& -e\int d\tau \mathbf A(\tau)\,  \exp
  \left\{
  -i\omega'
  \left(
    n'\cdot u_0 \tau
   +\frac{a^2}{n\cdot u_0} \int \limits^\tau d\tau' g^2(\tau')
  \right)
\right\},
\end{eqnarray}
where the dependence on proper time $\tau$ in the exponent is strongly nonlinear due to non-vanishing $a \sim 1$ and the presence of the smooth pulse envelope $g$.

\begin{figure}
  \includegraphics[scale=0.38,angle=-90]{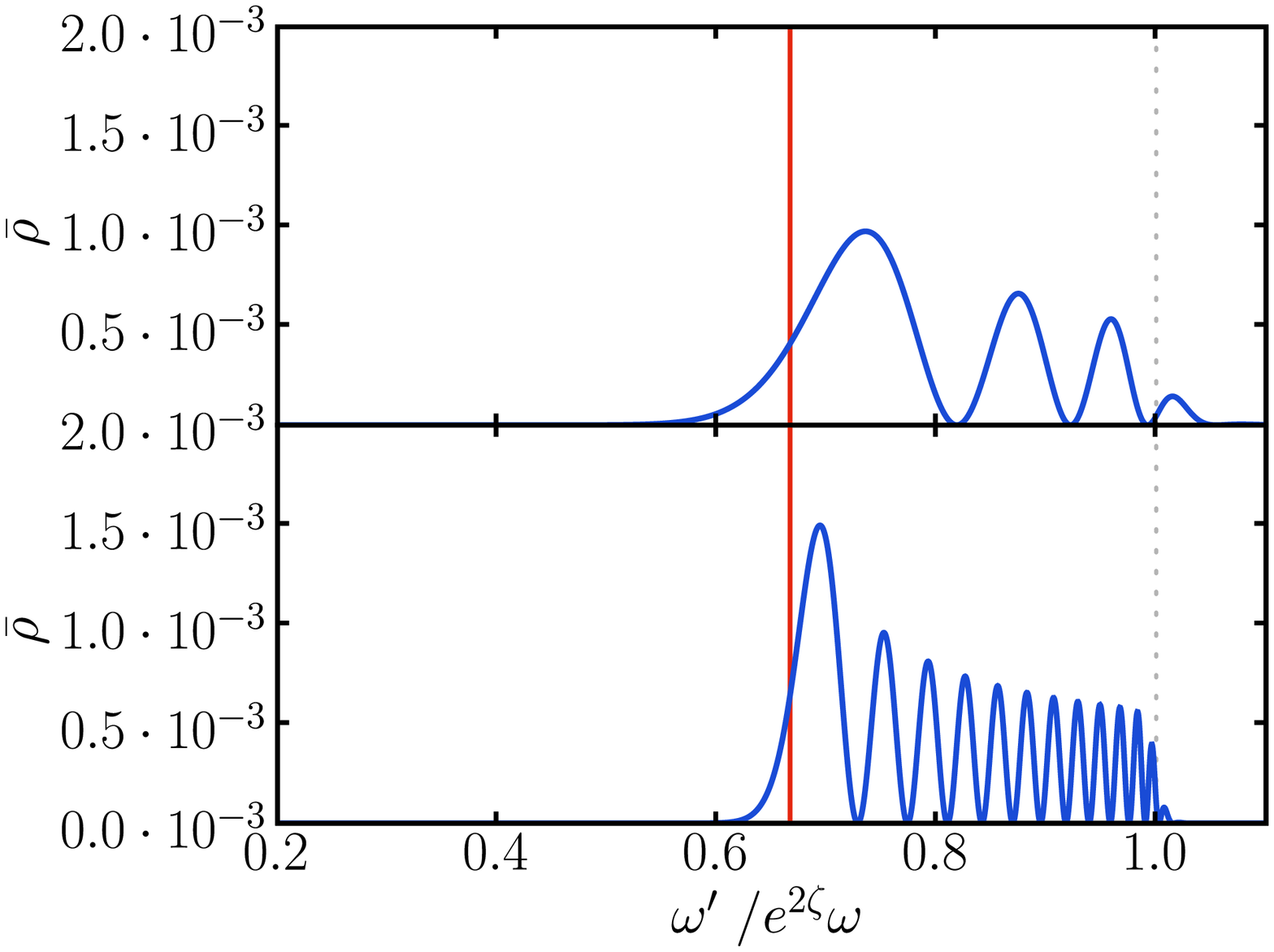}
  \includegraphics[scale=0.38,angle=-90]{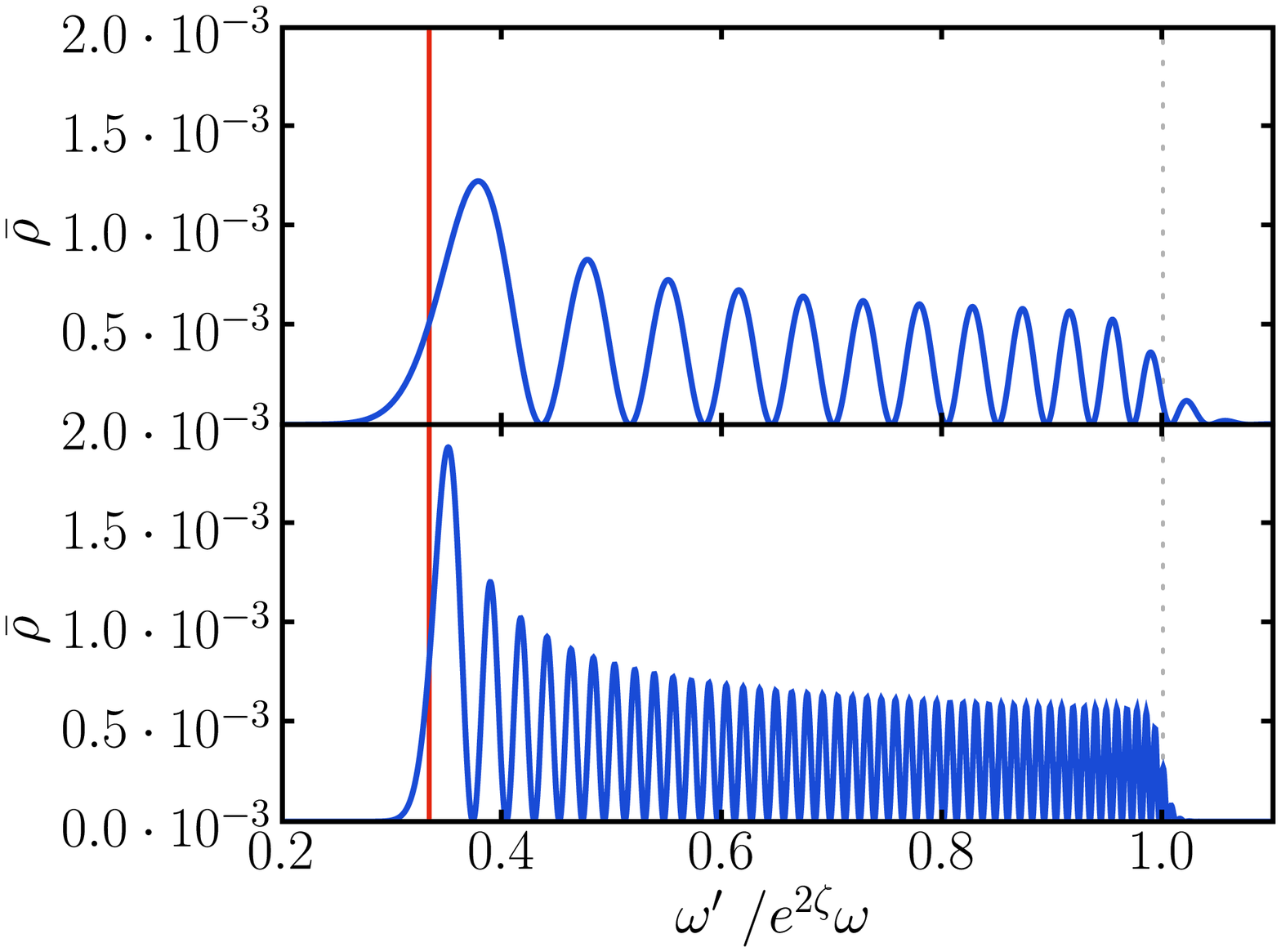}
  \caption{Normalised spectral density $\bar{\rho} = \rho/\tau_0$ as a function of normalised frequency $\omega'/e^{2\zeta}\omega$ for $a = 0.7$ (left panel) and $a = 1.4$ (right panel). Upper (lower) panels correspond to a laser pulse duration of $T_{0} = 25$ ($T_{0} = 100$ fs).  The full vertical (red) lines denote the nonlinear Compton edges for the ideal case of an infinite plane wave laser field, while dotted (grey) lines represent the respective linear Compton edges. For pulsed fields the spectral density covers the whole range between the two Compton edges.}
 \label{fig.backscattering}
\end{figure}

As a \textit{numerical} example, we have chosen the pulse shape (\ref{SECH}) advocated for in \cite{McDonald:1997} and determined the radiation spectrum for laser amplitudes $a=0.7$ and $a=1.4$ and pulse durations\footnote{The pulse duration $T_{0}$ in the \textit{lab} frame is defined by setting $\int d(k\cdot x) \, g_\sigma (k\cdot x) \equiv \omega T_0 \, {\rm max} \, (g_\sigma)$. For the solitonic pulse (\ref{SECH}), $T_0$ is related to the proper time pulse duration $\tau_0$ via $T_{\rm 0} = \pi e^{\zeta} \tau_0 $ .} $T_{\rm 0} = 25\ \rm fs$ and $100 \ \rm fs$ assuming circular polarisation as before. In Fig.~\ref{fig.backscattering} we have plotted the normalised spectral densities $\bar \rho \equiv \rho /T_{0}$ as a function of the normalised frequency $\omega^\prime / e^{2\zeta} \omega$.
Looking at the spectra one observes $N_\tau$ sub-peaks within the first harmonic signal resulting from the nonlinear $\tau$-dependent modulation in the exponent of (\ref{J.K.SMOOTH}). Radiation generated at different times $\tau$, and therefore at different effective laser intensities $a^2g^2(\tau)$
with effective Compton edges $\omega'/e^{2\zeta}\omega=(1+a^2g^2(\tau))^{-1}$, interferes thus generating the pattern of sub-peaks seen in Fig.~\ref{fig.backscattering}. There are important differences between smooth pulses and the interference for box-shaped flat top pulses which were discussed in~(\ref{JN.FINITE.N}) and below. Here, the interference pattern is due to the nonlinearity of the process ($a\sim 1$) together with the non-trivial envelope function $g$ of the laser and it disappears in the limit $a \to 0$. The smooth pulse sub-peaks are a sign of chirp in the emitted X-ray pulse. This however, is not the case for box-shaped pulses, where the side-peaks just reflect the discontinuous pulse shape and resemble the interference pattern of a single slit. These differences will be further investigated in Appendix \ref{app.time.structure} where the time structure of the scattered radiation is discussed. Summarising we conclude that sub-peaks in the harmonic signals are a purely nonlinear effect.

It has been known for a while that temporal modulations strongly affect the spectral densities, with the main effect being the additional oscillatory substructures \cite{Hartemann:1996zza,Krafft:2004,Debus:2009}. Our analysis above gives a rather simple explanation of this phenomenon. Interestingly, a similar pattern (with a similar explanation) has been observed recently for the rates associated with laser induced pair creation \cite{Hebenstreit:2009km} which is obtained from Thomson/Compton scattering via crossing.

We have found that the number $N_\tau$ of subsidiary peaks within the first harmonic scales linearly with the pulse duration $T_{0}$ and the intensity $a^2$ according to the empirical formula
\begin{eqnarray} \label{PULSE.NUMBER}
  N_\tau = 0.24 \, T_{0} [{\rm fs}] \, a^2 \; .
\end{eqnarray}
Reversing the arguments leading to (\ref{PULSE.NUMBER}) suggests the interesting possibility to actually determine the intensity of a laser pulse by counting the number of sub-peaks within the first harmonic if the pulse duration is known.

To the best of our knowledge, the harmonic substructure has never been observed experimentally. This may be due to the fact that most certainly the sub-peaks will get smeared out by a variety of mechanisms. These include effects such as (i) the ponderomotive force in focused beams, (ii) the influence of transverse intensity profiles as well as (iii) contributions due to the phase space distribution of the electron beam, in particular its energy spread $\Delta \gamma/\gamma$ and transverse beam emittance $\varepsilon$.
These effects will be studied in detail in the next subsections.
In particular, we will show how these effects can be minimised to allow for a possible experimental verification of the individual subpeaks with a $100$ TW class laser system.

\subsection{Spatial intensity profile}
\label{sect.spatial.profile}
Let us briefly turn to the effects associated with a transverse intensity profile of the laser beam. This is taken into account by choosing the complex vector potential
\be
  \vcb{A} = a \vc{\epsilon} \Psi({b},z) \, g_\sigma(k \cdot x) \, e^{i k \cdot x} \; ,
\ee
with a (linear or circular) polarisation vector\footnote{
Note that the use of unnormalised polarisation vectors $\{\mathbf e_+,\mathbf e_-\}$ is required here to be consistent with the normalisation of $A^\mu$ in (\ref{CIRC.P.W.}).}
\begin{eqnarray} \label{DEF.POL}
 \boldsymbol \epsilon \in
\left\{
\begin{array}{lll}
 \{\mathbf e_x,\mathbf e_y\} &\quad & \mbox{for linear polarisation}, \\
  \{\mathbf e_+,\mathbf e_-\}=
  \{{\mathbf e_x+i\mathbf e_y}, {\mathbf e_x-i\mathbf e_y}  \} &\quad & \mbox{for circular polarisation}.
\end{array}
\right.
\end{eqnarray}
and the transverse distance ${b} \equiv (x^2 + y^2)^{1/2}$. The paraxial approximation of the wave equation then yields the following equation for the transverse profile $\Psi$ \cite{McDonald:1997},
\begin{eqnarray}
  \triangle_\perp \Psi - 2i\omega(\partial_z \Psi) \left[ 1 - i \frac{g^\prime_\sigma}{g_\sigma} \right] = 0,
\label{eq.paraxial}
\end{eqnarray}
with the transverse Laplacian $ \triangle_\perp = \partial_x^2 +\partial_y^2$ and primes denoting derivatives with respect to $k \cdot x$. The standard Gaussian beam solution (with $g_\sigma = 1$) reads \cite{Chang}
\begin{eqnarray}
  \Psi_0(\rho,z) = \frac{w_0}{w(z)} \, \exp \left\{ -\frac{{b}^2}{w(z)^2} \right\} \, \exp \left\{ i \arctan \frac{z}{z_R} - i \frac{\omega
  {b}^2}{2R(z)} \right\} \label{eq.gaussian.profile}
\end{eqnarray}
with focal spot radius (``waist'') $w_0$, Rayleigh length $z_R = w_0^2 \omega/2$, beam radius $w(z) = w_0 (1 + z^2/z_R^2)^{1/2}$ and the curvature of the wave fronts $R(z) = z (1+ z_R^2/z^2)$.

In order for the Gaussian profile $\Psi_0$ to be an approximate solution of the paraxial equation~(\ref{eq.paraxial}), the temporal profile function $g_\sigma (k \cdot x) = g(\tau/\tau_0)$ has to satisfy $|\dot{g}/g| \ll \Omega$ \cite{McDonald:1997}. The  Gaussian $g (\tau) = \exp(- \tau^2/2\tau_0^2)$ does not have this property, since $|\dot{g}/g| \sim \tau$ is unbounded. For a hyperbolic secant, $g (\tau) = \sech (\tau/\tau_0)$, the condition can be fulfilled, as $|\dot{g}/g| = \tanh(\tau/\tau_0)/\tau_0$ is bounded \cite{Davis:1979zz,McDonald:1997}.

For a strongly focused beam such that $\Delta \equiv w_0/z_R = O(1)$, the intensity profile $\Psi_0$ will have to be corrected, since the fields derived from $\Psi_0$ solve Maxwell's equations up to terms of $O(\Delta^2)$~\cite{Davis:1979zz,Narozhny:2002}. Although these corrections to $\Psi_0$ are crucially important in some cases \cite{Salamin:2002gh,Bulanov:2004de}, we did not find them relevant for the electron trajectories or the emitted photon spectrum associated with the (almost) head-on collisions studied in this paper.

\begin{figure}[ht]
  \includegraphics[scale=0.37,angle=-90]{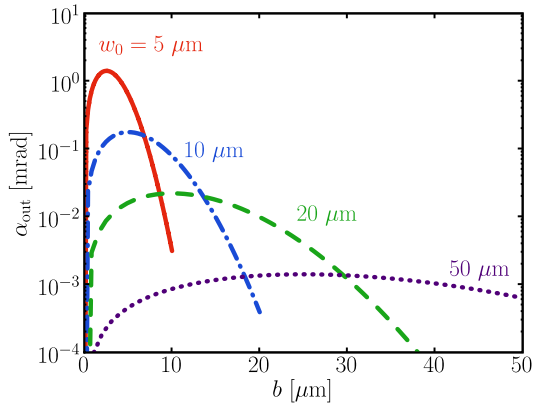}
  \caption{Effect of the ponderomotive force deflecting an electron with impact parameter
  ${b}$, initially in perfect alignment with the laser pulse (zero injection angle), after collision ejected with an angle $\alpha_{\rm out}$ with respect to the beam axis. Laser parameters: energy $3\,\rm J$, pulse duration $T_0 = 20\,\rm fs$, different focal radii $w_0$.}
\label{fig.ponderomotive}
\end{figure}

An important effect of the transverse beam profile is the ponderomotive force
$\mathbf F_p = - \boldsymbol \nabla m\mathbf A^2/2$ pushing the electrons away from regions of high intensity as they gain transverse momentum~\cite{McDonald:1986zz}. Accordingly, electrons with $\boldsymbol \beta \parallel \mathbf e_z$ before the scattering will leave the interaction region under an angle $\alpha_\text{out}$ with respect to the $z$-axis.
For fixed total pulse energy $W_{\rm tot}$ and pulse length $T_0$, the magnitude of the ponderomotive
force scales as $F_p \propto W_{\rm tot} / (T_0w_0^3) \propto {\rm max}\, (\alpha_{\rm out})$. This means
that the ponderomotive force leads to measurable effects only for very small waist size $w_0$.
To quantify the effect we simulate electron trajectories corresponding to head-on collisions assuming a laser pulse of $3$ J energy and $T_0=20\ \rm fs$ for different impact parameters ${b}$ (cf.\ Fig.~\ref{fig.ponderomotive}). For $w_0 = 5\ \rm \mu m$ the maximum deflection angle for $40\ \rm MeV$ electrons is about $1\ \rm mrad$ and for $10\ \rm \mu m$ it is roughly one order of magnitude lower. This is small compared to the typical angular scale of the emitted radiation which is of order $1/\gamma \sim 12\ \rm mrad$.

\begin{figure}[!ht]
\includegraphics[scale=0.37,angle=-90]{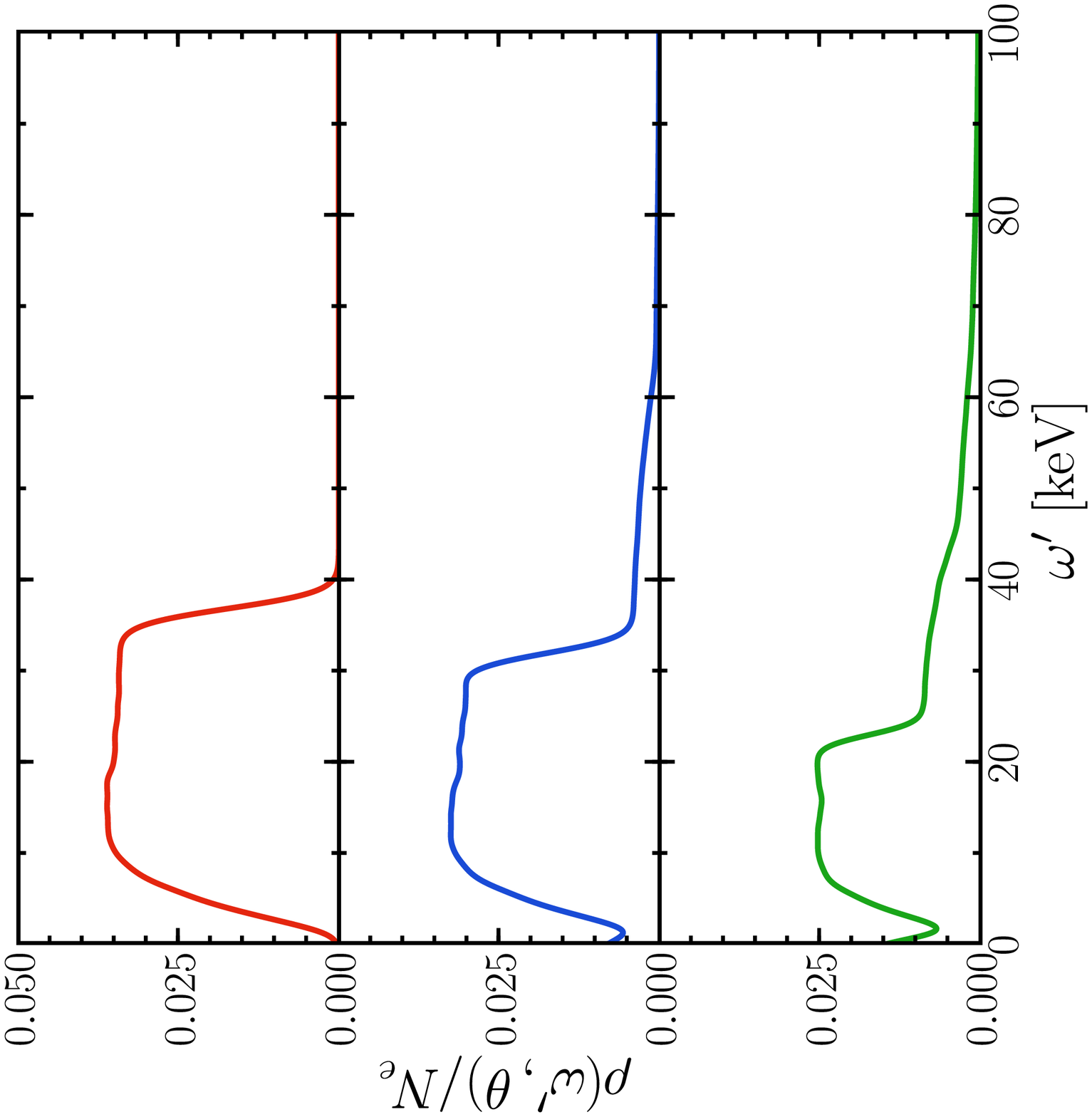}
\includegraphics[scale=0.37,angle=-90]{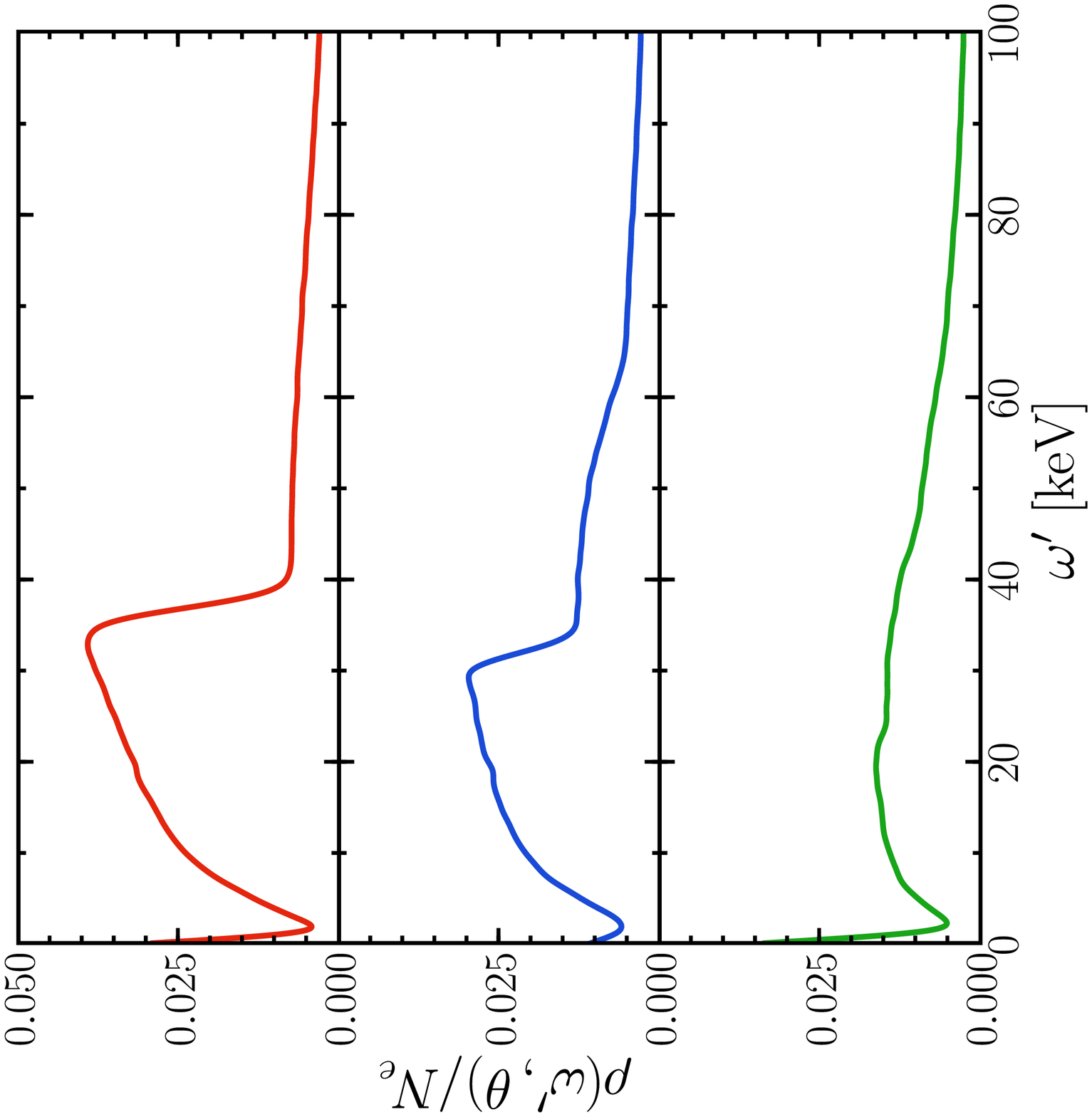}
\caption{Normalised spectral density $\rho(\omega',\theta)/N_e$ of the scattered radiation as a function of frequency $\omega^\prime$ and different scattering angles, $\theta=0,5,10$ mrad (from top to bottom). Assumes scenario: head-on collision of a dilute electron bunch ($N_e = 10000$) with a strongly focused ($w_0 = 5\, \rm \mu m$) laser pulse ($3$ J, $20$ fs). Left panel: circular polarisation. Right panel: linear polarisation.}
\label{fig.sim.1}
\end{figure}

Let us move on to our numerical simulations, choosing a $3$ J laser pulse described by either a circularly or linearly polarised Gaussian beam of pulse duration $T_0=20$ fs and wavelength $800\ \rm nm$ colliding head-on with a dilute electron beam. The spectral density for $N_e$ electrons is calculated as an incoherent superposition of the individual emission rates $\rho_i$ according to
\begin{eqnarray} \label{RHO.SUM}
 \rho(\omega') = \sum_{i}^{N_e} \rho_i (\omega') \; .
\end{eqnarray}
Figure~\ref{fig.sim.1} shows the spectrum $\rho(\omega',\theta)$ in the plane $\phi=0$ for a tightly focused laser with $w_0= 5\ \rm \mu m$ and a corresponding Rayleigh length of $z_R = 100\ \rm \mu m$. The peak values of the normalised amplitude, $a = 8.66$ and $a=6.12$ for linear and circular polarisation, respectively, are clearly in the nonlinear regime, the associated nonlinear Compton edge being approximately 1 keV,  cf.\ (\ref{OMEGAPRIME}). The electron bunch --- in this subsection we use cold electron bunches --- is modeled by a Gaussian with a bunch length of $1\ \rm ps$ and a transverse beam size of $r_b = 5\ \rm \mu m$. The spectral density, normalised to one electron, is shown as a function of the energy $\omega^\prime$ for different scattering angles, $\theta= 0,5$ and $10$ mrad. Due to the strong field gradients of the laser in both transverse and longitudinal directions, the spectrum is extremely broad for the parameter values chosen. The individual harmonics are not visible as the individual spectral lines are overlapping. It seems fair to describe the scattered photons as a broad continuum.

\begin{figure}[!ht]
\includegraphics[scale=0.37,angle=-90]{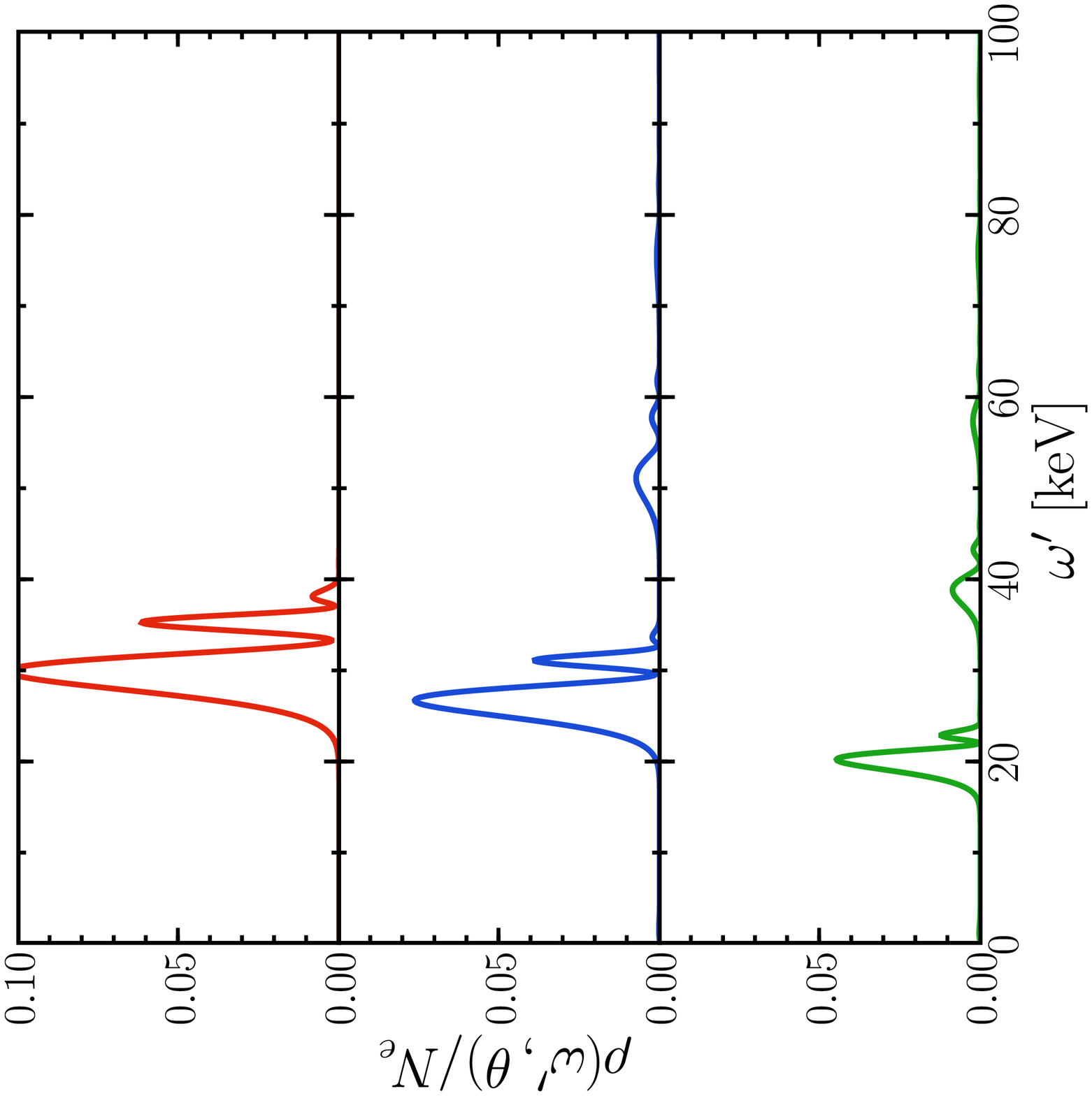}
\includegraphics[scale=0.37,angle=-90]{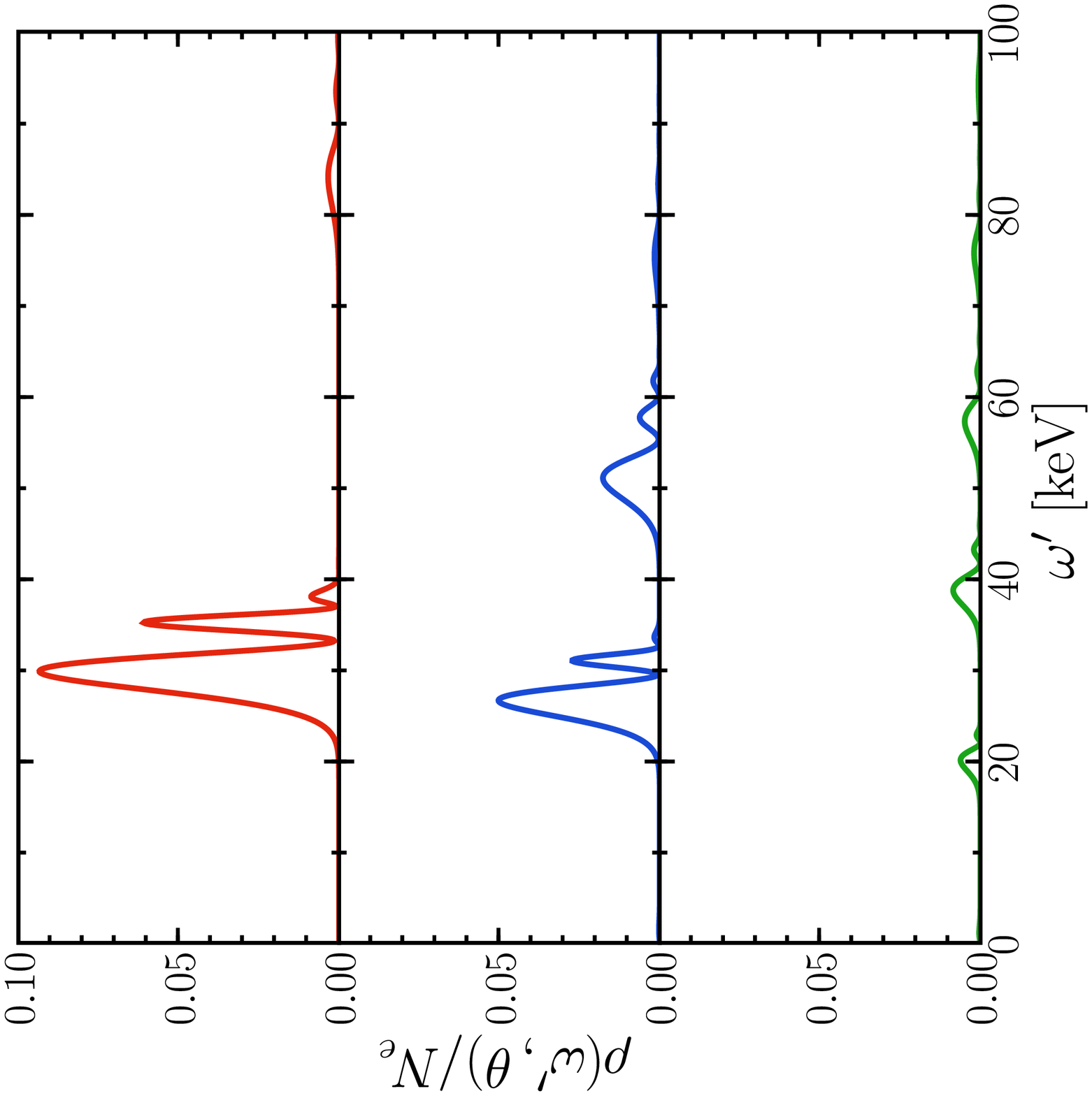}
\caption{Same as Fig.~\ref{fig.sim.1} but for $w_0 = 50\ \rm \mu m$.} \label{fig.sim.3}
\end{figure}

In comparison, Fig.~\ref{fig.sim.3} shows the backscattered spectrum for the same pulse ($3$ J, $T_0=20$ fs), but with a larger focal radius of $w_0 = 50\ \rm \mu m$ corresponding to a Rayleigh length of $z_R =10\ \rm mm$. The maximum value of the normalised amplitude is $a = 0.866$ ($a=0.612$) for linear (circular) polarisation, which corresponds to a nonlinear Compton edge of $27.5\ \rm keV$. The electron beam has the same transverse beam size $r_b$ and bunch length as before (Fig.~\ref{fig.sim.1}), so that the electron bunch exclusively probes the very centre of the focus where the laser intensity is almost constant. In fact, at $r_b= 5\,{\rm \mu m} = w_0/10$ the field intensity is only $1\%$ lower than at the centre of the focus. In contrast to Fig.~\ref{fig.sim.1}, the harmonics are now well-separated and clearly visible.

\subsection{A Scaling Law for the Spectral Density}
\label{sect.scaling.law}

For the following we assume that the laser waist size is much larger than the electron beam radius, $w_0 \gg r_b$, so that the electron beam interacts only with the centre of the laser focus. In this case the laser may be reasonably described by a pulsed plane wave in the relevant interaction region~\cite{Tomassini}.

\begin{figure}[!ht]
  \includegraphics[scale=0.55]{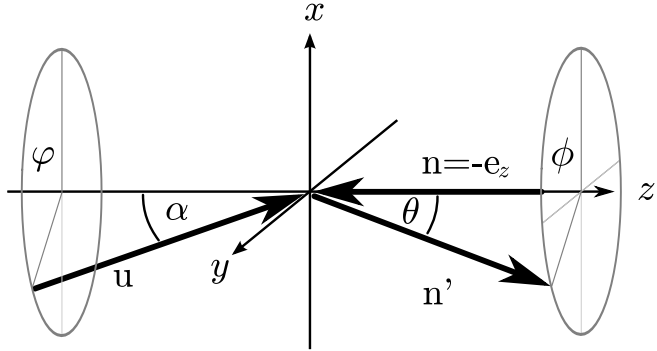}
  \caption{Sketch of beam geometry. The initial electron velocity is denoted by $\mathbf u$, while $\mathbf n$ ($\mathbf n'$) is the direction of the incoming (outgoing) photon.}
 \label{fig.beam.geometry}
\end{figure}

We may thus concentrate our attention on energy spread and emittance.\footnote{In fact, things may be turned around by using Thomson scattering as a diagnostic tool to measure electron beam parameters \cite{Leemans}.} According to \cite{Hartemann:2002} the former may be modelled by a distribution of the Lorentz factor $\gamma$ of width $\Delta \gamma$, centred at $\gamma_0$,  while the latter measures the transverse phase space volume of the beam via the correlator
\be \label{EMITTANCE}
  \varepsilon_{x} = \gamma_0 \beta_0 \sqrt{ \langle x^2 \rangle \langle \xi^2 \rangle - \langle x \xi \rangle^2} \simeq \gamma_0 \beta_0 \, r_b \, \Delta \alpha
  \; .
\ee
The expectation values $\langle \cdots \rangle$ refer to the transverse phase space distribution of the electron ensemble, usually taken to be Gaussian as well. In addition, we have defined the normalised transverse momentum, $\xi = |p_x/p_z| = \tan \alpha \approx \alpha$, which basically coincides with the injection angle $\alpha$ with respect to the beam axis  (chosen as the $z$ direction).  Obviously, while our previous considerations were assuming a head-on collisions ($\vc{\beta} \cdot \vc{n}= - \beta$), we now have to consider sideways injection allowing for a small angle $\alpha$ relative to the laser beam axis (see Fig.~\ref{fig.beam.geometry}), i.e.\ $\vc{\beta} \cdot \vc{n}= - \beta \cos\alpha$.
In this case we have to use the general relation (\ref{OMEGAPRIME}) for the scattered frequency depending on three scalar products,
\bea
  n \cdot n' &=& 1 + \cos \theta \; , \\
  n \cdot u &=& \gamma (1 + \beta \cos \alpha) \; , \\
  n' \cdot u &=& \gamma (1 - \beta \cos \alpha') \; .
\eea
Here, $\theta$ is the scattering angle between $\vcb{k}'$ and $-\vcb{k}$, $\alpha$ is the angle of incidence between laser and electron beams ($\vcb{p}$ and $- \vcb{k}$) and $\alpha'$ the angle between $\vcb{p}$ and $\vcb{k}'$. For a head-on collision, $\alpha = 0$ and $\alpha' = \theta$. Allowing for the possibility of linear polarisation we choose $\vc{\epsilon}_1 = \vcb{e}_x$ so that laser direction and polarisation define the $xz$ plane. This introduces another azimuthal angle, $\varphi$, for the electron momentum, $\vcb{p} = \gamma m \vc{\beta}$. We thus have in general $\vc{\beta} = \beta (\sin \alpha \cos \varphi, \sin \alpha \sin \varphi ,  \cos \alpha)$, $\vcb{n}' = (\sin \theta \cos \phi, \sin \theta \sin \phi,  \cos \theta)$ and
\be
\cos \alpha' = \cos(\theta-\alpha) - [\cos(\theta-\alpha) - \cos(\theta+\alpha)]  \sin^2 \frac{\phi-\varphi}{2}.
\ee
As we now have a distinguished vector transverse to the beam, namely $\vcb{p}_\perp \equiv (p_x, p_y)$, we are breaking axial symmetry and thus expect the radiation to develop an azimuthal dependence on $\phi$. The same is known to happen for linear polarisation (even for head-on collisions, $\alpha = 0$ \cite{Esarey:1993zz}) and has been used to detect higher harmonics \cite{Chen:1998}.

To proceed we need the dependence of the spectral density (\ref{RHO.DEF}) on the initial conditions, in particular the initial 4-velocity $u_0$. Clearly, this enters via the current (\ref{J.MU.K}) according to
\be \label{J.MU.K.U0}
  j^\mu (k') = e \int d\tau \, u^\mu (\tau; u_0) e^{-i k' \cdot x(\tau; u_0)} \; .
\ee
Writing $u_0 = \gamma_0 (1, \vc{\beta}_0)$ with $\vc{\beta}_0 = \beta_0 (\sin \alpha_0 \cos \varphi_0, \sin \alpha_0 \sin \varphi_0 ,  \cos \alpha_0)$ we see that the spectral density will depend on the initial electron energy $\gamma_0$ (via $\beta_0$) and the angles of incidence, $\alpha_0$ and $\varphi_0$, $\rho = \rho(\omega' , \vcb{n}' ; \gamma_0,  \alpha_0, \varphi_0)$.

How will this change with initial conditions? To address this question we view such a change as being due to a Lorentz transformation, $\Lambda: \, u_0 \to u_\Lambda = \Lambda u_0$,  where $\Lambda$ is defined in Appendix~\ref{APP1}. Under this assumption one can derive a scaling formula that relates the spectral density for the transformed initial conditions to the original one. Choosing the initial kinematics of a head-on collision ($\alpha_0 = 0$, $\varphi_0=0$) for simplicity and $u_\Lambda = u_\Lambda (\gamma, \alpha, \varphi)$ we find
\be \label{SCALING.LAW}
  \rho(\omega' , \vcb{n}' ; \gamma,  \alpha, \varphi) = M (\gamma, \alpha, \varphi; \gamma_0 , 0 , 0) \, \rho({\omega}'/h , \vcb{n}' ; \gamma_0,  0,0) \; .
\ee
Here, we have defined a rescaling factor for the scattered frequency
\be
     h \equiv \frac{\nu'(u_\Lambda)}{\nu'(u_0)}
\ee
with $\nu'(u)$ as in (\ref{NUPRIME.CLASSICAL}) and a `transition function' which depends only
on kinematic quantities
\be \label{eq.def.M}
  M = \frac{\nu'(u_\Lambda)}{\nu'(u_0)} \, \frac{(n \cdot u_0)^2}{(n \cdot u_\Lambda)^2} \times
  \left\{ \begin{array}{ll}
  1-4s + 4 s^2 \varsigma   & \quad \rm linear\ polarisation \\
  1 - 2s + 2s^2  & \quad \rm circular\ polarisation
\end{array}
\right. \; , \ee
where we have introduced
\be \label{S}
  s \equiv \frac{h{\omega}'}{\omega} \,  \frac{|\epsilon \cdot u_\Lambda|^2}{(n \cdot u_\Lambda)^2} \;
\quad
 \varsigma \equiv \frac{ (\mathbf u_\Lambda)_\perp^2}{|\epsilon \cdot u_\Lambda|^2} .
\ee
Here, $\epsilon=(0,\boldsymbol \epsilon)$ denotes a single, real or complex, polarisation vector describing linear or circular polarisation (\ref{DEF.POL}). Averaging $M$ corresponding to linear polarisation over the azimuthal angle $\varphi$ one naturally recovers $M$ for circular polarisation. The first factor in (\ref{eq.def.M}) is related to the Jacobian of a Lorentz transformation for going from head-on to side-injection geometry and accounts for the fact that the radiation is peaked in the forward direction of the electron. The derivation of the scaling law (\ref{SCALING.LAW}) is deferred to Appendix~\ref{APP1}.

The main virtue of the scaling property (\ref{SCALING.LAW}) obeyed by the spectral density $\rho$ is its use in calculating what is called the `warm spectral density' $\rho_W (\omega')$. The latter is defined as the expectation value of the (`cold') spectral density $\rho(\gamma, \alpha,\varphi)$ taken in the initial ensemble of $N_e$ electrons, characterised by the normalised distribution $f(\gamma, \alpha,\varphi)$ of initial energies and injection angles. Thus we have
\begin{eqnarray}
  \rho_W(\omega') &=& N_e \int d\gamma d\alpha d\varphi \,  f(\gamma,\alpha,\varphi) \, \rho(\omega' ; \gamma, \alpha,\varphi) \nonumber \\
  &=& N_e \int d\gamma d\alpha d\varphi \,  f(\gamma,\alpha,\varphi) \, M (\gamma, \alpha ,\varphi; \gamma_0,0,0) \, \rho (\omega^\prime/h; \gamma_0, 0,0) \; .
  \label{eq.def.warm_spectral_density}
\end{eqnarray}
As pointed out in \cite{Hartemann:2002}, in this way one avoids to perform a summation over an ensemble of test-particles of the form (\ref{RHO.SUM}).
The distribution function  $f(\gamma,\alpha,\varphi)$ in (\ref{eq.def.warm_spectral_density}) is taken to be a product of a Gaussian in $\gamma$, a $\chi$-distribution (with $2$ degrees of freedom) in $\alpha$ and a uniform distribution in $\varphi$,
\be \label{DISTRIBUTION}
  f(\gamma,\alpha,\varphi) = \frac{1}{2\pi}f_\gamma(\gamma) f_\alpha (\alpha)  = \frac{1}{(2\pi)^{3/2} \Delta \gamma} \exp \left\{ -\frac{(\gamma-\gamma_0)^2}{2(\Delta
  \gamma)^2} \right\}  \frac{\alpha}{(\Delta \alpha)^2} \exp \left\{ -\frac{\alpha^2}{2(\Delta \alpha)^2} \right\}
\ee
as we assume the electron beam to be axial symmetric with both transverse components $u_{x0}$ and $u_{y0}$ normally distributed and uncorrelated.
\begin{figure}[!ht]
 \includegraphics[scale=0.40,angle=-90]{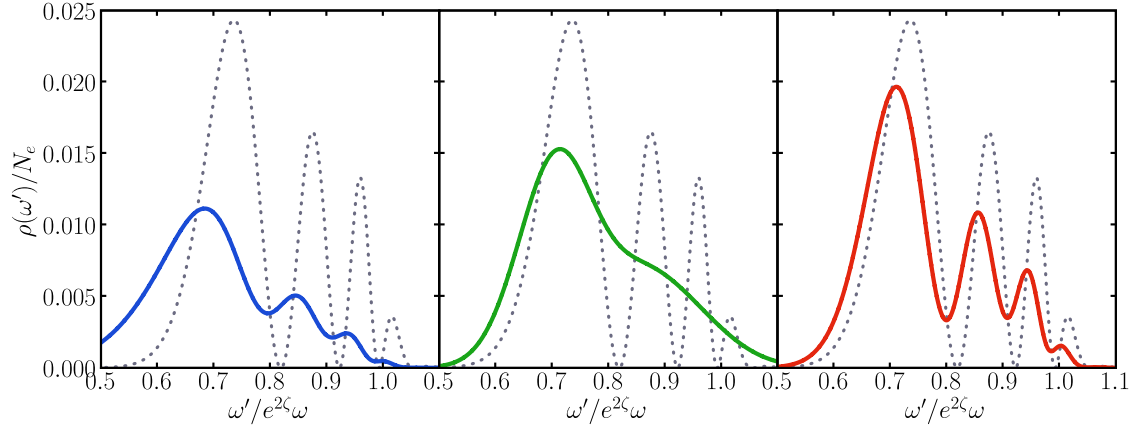}
  \caption{
The warm spectral density for ELBE~\cite{ELBE} parameters (left panel), LWFA electrons~\cite{Osterhoff:2008zz} (centre panel) and for our proposed parameters (right panel, see Tab.~\ref{tab.parameters}). For comparison the cold spectral density is shown also in each plot as grey dotted curve. The smearing of the harmonic sub-peaks due to finite emittance and energy spread is clearly visible. Very low values for both the energy and angular spread are needed for a possible observation of the sub-peaks.
}
\label{fig.scaling}
\end{figure}
In order to resolve the substructures in an experiment, one has to  fine-tune several parameters. First of all it seems sensible to only have a small number of subpeaks (say less then five), so that they can be clearly resolved. This is achieved by adopting values of $a = 0.7$ and $T_0 = 25 \ \rm fs$ (these values correspond to the upper left panel of Fig.~\ref{fig.backscattering}). It is important to have good control of both the energy spread and the emittance. The energy spread tends to smear out the small subpeaks, in particular at the high energy end of the spectrum, whereas emittance affects the whole range.

A common source for ultrarelativistic electron beams are linacs like the ELBE accelerator at the FZD~\cite{ELBE}. It is capable of producing electron bunches with a very low energy spread of $\Delta \gamma/\gamma_0=10^{-3}$ and transverse emittance of about $\varepsilon_x=1.5$ mm mrad. On the other hand, new laser based acceleration schemes like laser wake field acceleration (LWFA) report the production of electron bunches with $\Delta \gamma/\gamma_0 = 3.5\%$ and angular divergence of $\Delta \alpha = 0.68$ mrad at $\gamma_0 = 400$ \cite{Osterhoff:2008zz}.
\begin{table}[ht]
\caption{Beam parameters required for observing the substructure in the fundamental harmonic of nonlinear Thomson scattering employing a $100$ TW laser.} \label{tab.parameters}
\begin{ruledtabular}
\begin{tabular}{llc}
parameter &  & proposed value \\
\hline
laser frequency & $\omega\,[\rm e V]$ & $1.5$ \\
laser amplitude & $a$ & $0.7$ \\
laser pulse length & $T_0\,[\rm fs]$ & $25$ \\
laser focal radius & $w_0\, [\rm \mu m]$ & $50$ \\
electron beam radius & $r_b\, [\rm \mu m]$ & $5$ \\
electron bunch length & $L_b$ & $\ll z_R = 10\, \rm mm$ \\
electron energy spread & $\Delta \gamma/\gamma_0$ & $0.001$ \\
electron transverse emittance & $\varepsilon_{x}\, [\rm mm\, mrad]$ & $0.7$ \\
\end{tabular}
\end{ruledtabular}
\end{table}
The numerical results for the warm spectral densities using both ELBE and LWFA electron beam parameters are shown in Fig.~\ref{fig.scaling} which also includes a comparison with the respective cold spectral densities. Clearly, the harmonic sub-peaks are smeared out for ELBE parameters and even more so for LWFA. For ELBE it is the emittance which is too high for the observation of the sub-peaks. A LWFA electron bunch, on the other hand, has too large an energy spread --- despite its low angular divergence. We conclude that in order to resolve the sub-peaks both energy spread and emittance need to be sufficiently small. In Tab.~\ref{tab.parameters} we list a suitable set of parameters which allow to observe the harmonic sub-peaks with a $100$ TW laser. For petawatt lasers, the strong constraint on the emittance may be relaxed because they are capable to achieve nonlinear peak intensities $a^2 \gtrsim 1$ across larger spot sizes $w_0$ such that larger electron beam radii $r_b$ can be tolerated. The essential quantity in (\ref{DISTRIBUTION}) is $\Delta \alpha \propto \varepsilon/r_b$ and not the emittance itself.

\section{Discussion and Summary}
\label{sect.summary}

Backscattering of an optical laser beam by relativistic electrons has become important as a tunable source of X-rays (see e.g.~\cite{Sardinien:2008}). At low laser intensities ($a_0^2 \ll 1$) the relevant physics is adequately described in terms of the conventional classical or QED treatment (Thomson or Compton scattering, respectively). At high intensities ($a_0^2 \gtrsim 1$), however,  one enters the relativistic nonlinear regime where multi-photon processes become important. These require a \textit{strong-field} QED approach adopting a Furry picture where Volkov electrons dressed by the external field replace the ordinary QED electrons. As a result, the backscattered photon spectrum deviates significantly from the one corresponding to the standard Klein-Nishina formula for (linear) Compton scattering. By the correspondence principle, one expects that large photon numbers should allow for a classical description of the laser beam.  Indeed, we find that classical Thomson scattering (i.e.\ bremsstrahlung by an accelerated charge in the external laser field) yields the same answer (e.g.\ for unpolarised cross sections) as strong-field QED at leading order in the small recoil parameter, $x_1$. For an optical $100$ TW laser and $40\,\rm MeV$ electrons, such as in operation at Forschungszentrum Dresden-Rossendorf, one has $ x_1 \simeq 10^{-3}\, \ell$, where $\ell$ is the number of laser photons involved. The sharp decrease of emission probabilities with photon number $\ell$ corroborates the validity of working in the Thomson limit. In particular, as we have shown, effects due to finite size and realistic shape of both laser and electron beams may be easily addressed within this framework.

Our main focus was the search for suitable conditions allowing to experimentally observe the signatures of the non-linear, multi-photon processes in question. In particular, we have found that the rich substructure in the spectral density of the first harmonic strongly depends on the combination of short pulses and high intensity. To quantitatively assess this dependence, we established a simple scaling law for the spectral density by means of which we could estimate the effects of electron emittance and energy spread on this substructure.

Turning to experimental prospects we conclude that the ELBE/DRACO constellation at Forschungszentrum Dresden-Rossendorf offers the possibility to clearly detect the red-shift of the Compton edge (as already argued in \cite{Harvey:2009ry}) as well as the generation of higher harmonics (i.e.\ their intensity distribution and angular dependence) as long as the laser pulse is not focussed too strongly ($a_0 = O(1)$) and of sufficiently short duration. Higher intensities, accomplished by strong focussing, lead to a near-continuous backscattered radiation spectrum reaching far beyond $100\, \rm keV$ in the ultraviolet. This spectrum emerges from the superposition of many higher harmonics, modified by the temporal and spatial variations of the laser pulse. This renders a clear-cut verification of high-intensity signals rather difficult.

In summary we have presented realistic results for the Thomson/Compton backscattering spectra of optical laser photons by relativistic electrons assuming an experimental set-up that can be realised at facilities already in operation. The subtle interplay between beam and intensity parameters may be fine-tuned in such a way that the observation of intensity effects due to the increased effective electron mass seems feasible for the first time.

\section{Acknowledgements}

The authors gratefully acknowledge stimulating discussions with T.~E.~Cowan, C.~Harvey, A.~Ilderton, K.~Langfeld, K.~Ledingham, R.~Sauerbrey, R.~Sch{\"u}tzhold, G.~Schaller, H.~Schwoerer, V.~G.~Serbo and A.~Wipf.

\appendix
\section{Derivation of the Scaling Law}
\label{APP1}

The scaling law (\ref{SCALING.LAW}) relates the spectral densities $\rho$ for different geometries, in particular for different initial electron velocities, $u_0$. It is useful to describe such a change in geometry as being due to a Lorentz transformation, $\Lambda: u_0^\mu \to u_\Lambda^\mu = \Lambda^\mu_{\;\;\nu} u^\nu$, where $\Lambda$ is a Lorentz transformation composed of transverse rotations $\Lambda_R$ and a boost $\Lambda_B$ which may be parameterised by
\begin{align}
  \Lambda_R(\alpha,\varphi) &=
  \left(
  \begin{array}{cccc}
   1 & 0&0&0 \\
   0 & \cos \alpha + \sin^2 \varphi(1-\cos\alpha)
     & -\sin\varphi\cos\varphi(1-\cos\alpha) & \cos\varphi \sin\alpha \\
   0 & -\sin\varphi\cos\varphi(1-\cos\alpha)
     &\cos\alpha +\cos^2\varphi(1-\cos\alpha)& \sin\varphi \sin\alpha \\
    0& -\cos\varphi \sin \alpha & -\sin\varphi \sin\alpha & \cos\alpha
  \end{array}
  \right),\\
  \Lambda_B(\zeta) &= \left(
  \begin{array}{cccc}
   \cosh \zeta & 0&0&- \sinh \zeta \\
   0 & 1 & 0 &0 \\
   0 & 0&1&0\\
   -\sinh \zeta & 0 & 0 & \cosh \zeta
\end{array}
\right) \; ,
\end{align}
for a rotation with angle $\alpha$ around the axis $\mathbf v_\varphi = (-\sin\varphi,\cos \varphi,0)$ perpendicular to the $z$-axis and a boost along the $z$-axis with rapidity $\zeta = \cosh^{-1} \gamma$, changing the electron's energy, respectively. We construct $\Lambda$ as follows: First rotate $\mathbf u_0(\gamma_0,\alpha_0,\varphi_0)$ parallel to the $z$-axis, then apply the boost changing the electron's energy from $\gamma_0$ to $\gamma$ and finally rotate to the new direction $\alpha,\varphi$
\be
\Lambda = \Lambda_R(\alpha,\varphi)\Lambda_B(\gamma)\Lambda_B^{-1}(\gamma_0)\Lambda_R^{-1}(\alpha_0,\varphi_0).
\ee
According to (\ref{J.MU.K.U0}) the spectral density $\rho$ depends on the initial conditions through the orbit $x(\tau; u_0)$ and its velocity $u(\tau; u_0)$. In what follows, the direction of observation $n'$ is kept fixed. Let us consider the simplest case first, namely backscattering and a head-on collision, implying a change only in the electron energy,  $\gamma_0$ to $\gamma$ (as $\alpha = \varphi=0, \Lambda_R = 1$). In this case, the spectral density obeys the simple scaling relation
\begin{eqnarray}
 \rho(\omega',\gamma) = \rho(\omega'/h,\gamma_0) \label{eq.scaling.law.simple}
\end{eqnarray}
with a rescaled frequency $\omega'/h$ where $h = \nu'_1(u_\Lambda)/ \nu'_1(u_0)$ and $\nu'_1$ as in (\ref{NUPRIME.CLASSICAL}).

If we also allow for a change in the direction of $u_0$, the height of the spectral peak will certainly change, since the radiated intensity is peaked in the direction $\vc{\beta}$ of the electron. Thus, we have to use the modified ansatz $\rho(\omega'; u_\Lambda) = M(u_\Lambda, u_0) \rho(\omega'/h;u_0)$ with a transition function $M$ which we calculate in what follows, for low intensities ($a^2 \ll 1$) and arbitrary initial value $u_0^\mu$, but strictly staying within the backscattering geometry, i.e.\ $\mathbf n' = -\mathbf n$, $n' \cdot n = 2$. Linearising the orbit expressions (\ref{UMU}) and (\ref{XMU}) in the gauge field  $A$,
\begin{eqnarray}
 u^\mu(\tau;u_0) &=& u^\mu_0  - A^\mu + n^\mu \frac{A\cdot u_0}{n\cdot u_0}, \\
 x^\mu(\tau;u_0) &=& x_0^\mu +u_0^\mu \tau - \int \limits_0^\tau d\tau' A^\mu (\tau')
                     + n^\mu \int \limits_0^\tau d\tau' \frac{A(\tau')\cdot u_0}{n\cdot u_0} \; .
\end{eqnarray}
the three-vector part of the electron current (\ref{J.MU.K.U0}) becomes, to $O(A)$,
\begin{eqnarray} \label{J3}
\mathbf j(\omega')
&=& e\int d\tau \left[
  \frac{A\cdot u_0}{k\cdot u_0} \mathbf n - \mathbf A
  - 2 i \frac{\omega'}{n\cdot u_0} \mathbf u_0   \int \limits^\tau_0 d\tau' A(\tau')\cdot u_0
\right] e^{-i \omega' n'\cdot u_0 \tau} \; .
\end{eqnarray}
It is convenient to adopt a complex-valued vector potential $A^\mu = (0, \mathbf A)$ with $\mathbf A = a \boldsymbol \epsilon e^{i \Omega \tau} g(\tau/\tau_0)$) with $\Omega\tau = k \cdot x$ and the polarisation vectors $\boldsymbol \epsilon$ as in (\ref{DEF.POL}).

The inner integral in (\ref{J3}) yields, after an integration by parts,
\begin{eqnarray}
  \int \limits^\tau d\tau' A^\mu(\tau') &=& \frac{1}{i\omega n\cdot u_0}A^\mu(\tau)
  \left(1 + \mathcal O(1/\Omega\tau_0 ) \right) \; .
\end{eqnarray}
For sufficiently long pulses, $\Omega\tau_0 \gg 1$, the second term can be neglected. Within these approximations, the result for the electron current reads
\begin{eqnarray}
  \mathbf j(\omega') &=& -e a \int d\tau \,
     g(\tau) e^{-i(\omega'n'\cdot u_0 - \omega n\cdot u_0) \tau}\, \mathfrak j_0,
\end{eqnarray}
where
\begin{eqnarray}
\mathfrak j_0 =
     \frac{\boldsymbol \epsilon \cdot \mathbf u_0}{k\cdot u_0} \mathbf n
   + \boldsymbol \epsilon
   - 2 \frac{\omega' \boldsymbol \epsilon \cdot \mathbf u_0}{\omega (n\cdot u_0)^2}  \mathbf u_0.
\end{eqnarray}
Using (\ref{RHO.3D}) the spectral density becomes
\begin{eqnarray}
  \rho(\omega';u_0) = \frac{e^2a^2}{16\pi^3} \frac{\omega'}{(n\cdot u_0)^2} |\tilde g(\omega - \omega'(\nu/\nu'))|^2 |\mathbf n' \times \mathfrak j_0|^2 \;
\label{eq.scaling.law.spect.density}
\end{eqnarray}
with $\nu'$ from (\ref{NUPRIME.CLASSICAL.LINEAR}). Here and in the following $\tilde g$ denotes the Fourier transform of the envelope function $g$, and
\begin{eqnarray}
 |\mathbf n' \times \mathfrak j_0|^2 = |\boldsymbol \epsilon|^2
    + 4t^2 |\boldsymbol \epsilon \cdot \mathbf u_0|^2 (\mathbf u_0^2 -(\mathbf n\cdot \mathbf u_0)^2)
    - 4t   |\boldsymbol \epsilon \cdot \mathbf u_0|^2
\label{eq.n.times.j}
\end{eqnarray}
with $t=\omega'/\omega \, (n\cdot u_0)^2$.
The form of $\tilde g$ in (\ref{eq.scaling.law.spect.density}) lends support to the scaling behaviour of the frequency adopted in (\ref{eq.scaling.law.simple}). On changing the geometry, $u_0 \to u_\Lambda$, and replacing the frequency, $\omega' \to \bar \omega'= h \omega'$, with $h = \nu' (u_\Lambda)/\nu'(u_0)$,
the function $\tilde g$ remains invariant.

To determine the transition function $M$, we chose the special head-on geometry characterised by $u_0 = \gamma_0 (1,0,0,\beta_0)$ as a reference, and a second, different geometry characterised by $u_\Lambda = \Lambda u_0$ or, equivalently, the injection energy $\gamma$ and the injection angles $\alpha$ and $\varphi$ (cf.~Fig.~\ref{fig.beam.geometry}), which yields
\begin{eqnarray}
  M(u_\Lambda,u_0) = \frac{\rho(\bar \omega';u_\Lambda)}{\rho(\omega';u_0)}
  = \frac{\bar \omega'}{\omega'} \frac{(n\cdot u_0)^2}{(n\cdot u_\Lambda)^2}
  \frac{|\mathbf n' \times \mathfrak j_0(\gamma,\alpha,\varphi)|^2}{|\mathbf n' \times \mathfrak j_0(\gamma_0,0,0)|^2}.
\end{eqnarray}
Evaluating (\ref{eq.n.times.j}) with some explicit polarisation vectors one eventually arrives at (\ref{eq.def.M}) and (\ref{S}). To approximately include the non-linear case when $a \gtrsim 1$, we finally substitute the nonlinear scattered frequency (\ref{NUPRIME.CLASSICAL}) for $\nu'$ (\ref{NUPRIME.CLASSICAL.LINEAR}) in the definition of $h$, i.e.
\begin{eqnarray}
 \nu' \to \nu'_1 = \frac{n\cdot u_0 \nu }{n'\cdot u_0 + n'\cdot n \displaystyle \, \frac{a^2}{2n\cdot u_0}} \; .
\end{eqnarray}
In the nonlinear regime, the scaling law perfectly describes changes in the electron's initial energy. However, changes in the angles of incidence, in particular $\alpha$, are rendered accurate only for $\alpha \ll 1$. Nevertheless, for the purposes of this paper, the scaling law is sufficient to account for the typical angular divergence of the electron beams.

\section{Time Structure of the Backscattered pulse}
\label{app.time.structure}

To complete our analysis of effects due to finite pulse duration let us briefly comment on the temporal structure of the backscattered pulse. The Fourier transform $\mathbf F(t)$ of the electron current (\ref{J3}),
\begin{eqnarray}\label{F}
 \mathbf F(t) = \int \limits_{-\infty}^\infty d\omega' \, \mathbf j(\omega')\, e^{i\omega' t}
\end{eqnarray}
provides information on the time structure of the scattered pulse in the lab frame.
\begin{figure}
 \includegraphics[scale=0.35,angle=-90]{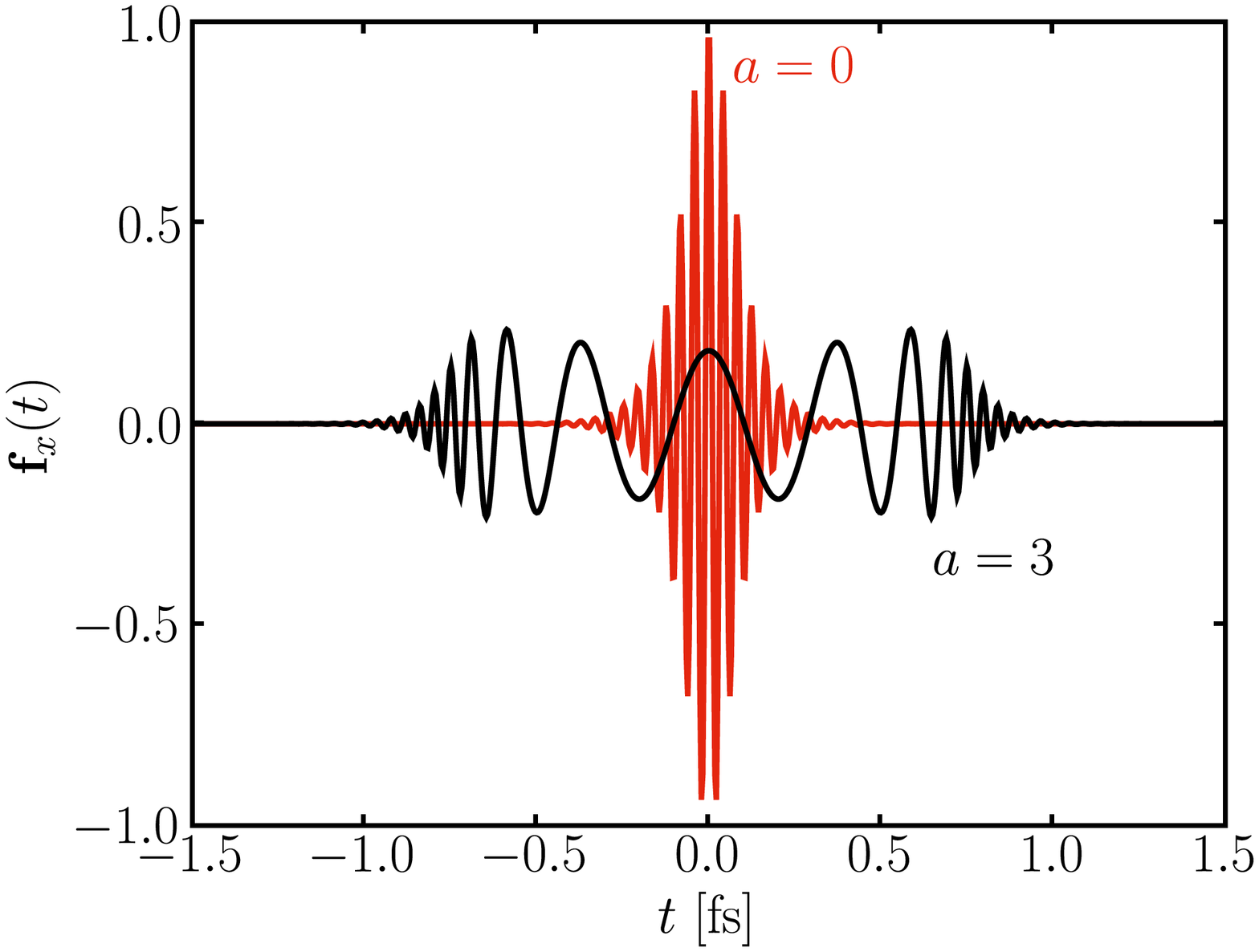}
 \includegraphics[scale=0.35,angle=-90]{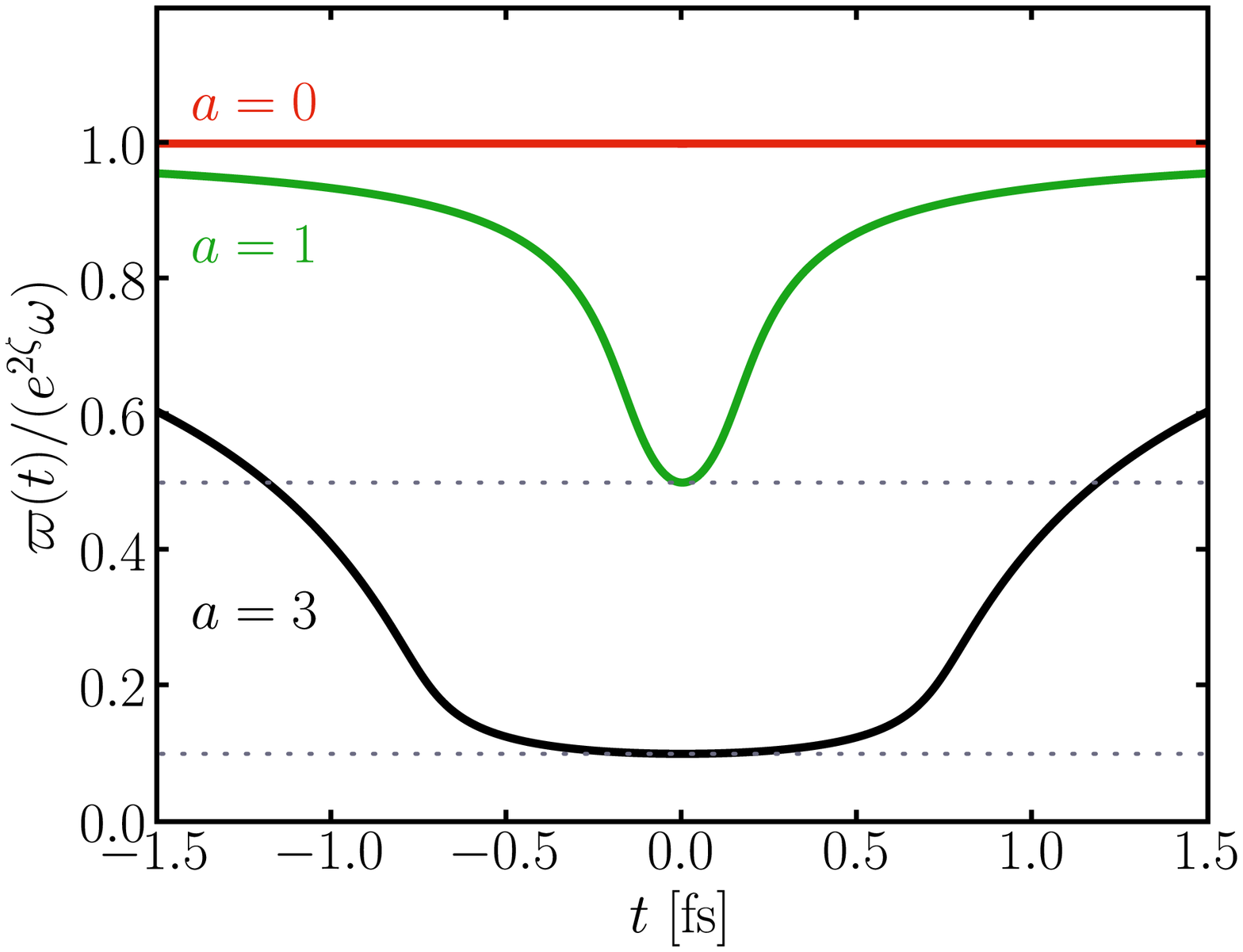}
\caption{Time structure of the scattered X-ray pulse for a circularly polarised laser with 'solitonic' pulse shape ($\omega=1\ \rm e V$, $\sigma=10\ \rm eV^{-1}$, $e^\zeta = 10$). Left panel: Electron current $\mathbf f$ defined in (\ref{eq.timedep2}) as a function of time $t$. Red (light) line: low intensity ($a \to 0$); black line: high intensity ($a=3$). Right panel: Normalised time dependent frequency $\varpi$ as a function of time $t$. Red (light) and black lines as before. Additional green curve: intermediate intensity ($a = 1$). Dotted lines: Thomson limit ($\ell = 1, \nu \to 0$) of scattered frequencies (\ref{NUPRIME.GENERAL}). } \label{fig.timestructure}
\end{figure}
If the envelope function $g$ in the current (\ref{J.K.SMOOTH}) is chosen as a 'solitonic' pulse as in (\ref{SECH}), the inner integral may be evaluated analytically with the result
\begin{eqnarray}
  \int \limits^\tau d\tau' g^2(\tau') &=& \tau_0 \tanh \left(\frac{\tau}{\tau_0}\right).
\end{eqnarray}
Thus, (\ref{F}) becomes
\begin{eqnarray}
 \mathbf F(t) &=& -e\int d\tau \mathbf A(\Omega\tau,\tau/\tau_0)\, \delta
 \left(
   t-\Big[ n'\cdot u_0\tau + \frac{a^2}{n\cdot u_0} \tau_0 \tanh \frac{\tau}{\tau_0} \Big]
 \right) \nonumber \\
 &\equiv&
 -ea \exp(\zeta) \; \mathbf f(t)
\label{eq.timedep1}
\end{eqnarray}
with
\begin{eqnarray}
   \mathbf f(t)
  &=& \frac{\mathbf A(\varpi(t) t,t/\sigma(t))}{a} \left( 1 + \frac{a^2}{\cosh^2 t/\sigma(t)} \right)^{-1}
\label{eq.timedep2},
\end{eqnarray}
where we defined a time dependent effective width $\sigma(t)$ via the transcendental equation
\be \label{def.sigma.t}
\frac{t}{\sigma(t)} + a^2 \tanh \frac{t}{\sigma(t)} = e^\zeta \frac{t}{\tau_0} \; ,
\ee
and a time dependent frequency $\varpi(t) \equiv \omega e^\zeta \tau_0/\sigma(t)$. By construction, the product of width and frequency is constant, $\varpi(t) \sigma(t) = \omega e^\zeta \tau_0$. The inversion of (\ref{def.sigma.t}) has to be done numerically. The results are shown in the left panel of Fig.~\ref{fig.timestructure} for two different values of $a$ and for $\omega = 1\ \rm eV$, $\tau_0 = 1 \rm eV^{-1}$, which corresponds to a pulse length of $T_{0} \approx 20\ \rm fs$ in the lab frame with $\exp(\zeta) = 10$. The main features are (i) an increase of the scattered pulse length for larger $a$ and (ii) a double chirp of the backscattered pulse due to the time dependent frequency $\varpi(t)$ -- the frequency decreases towards its minimum at the centre of the pulse and then increases.

The chirp in the backscattered signal is a combined effect of the non-linear interaction and the non-trivial envelope function $g$ with finite pulse length $\sigma$. Both features are required for the chirp to be present. Hence, if either $a\to 0$ or $\sigma \to \infty$ the chirp vanishes. For box-shaped, flat-top envelope functions, the time dependent frequency $\varpi(t)$ will be discontinuous at the beginning of the pulse, jumping from its linear value $e^{2\zeta}\omega$ to its nonlinear value, $e^{2\zeta}\omega/(1+a^2)$, and back again at the end of the pulse. In the intermediate regime $\varpi(t)$ will stay constant so that there will be no chirp for this particular case.

\end{document}